\documentclass{article}

\usepackage{arxiv}

\usepackage[utf8]{inputenc} 
\usepackage[T1]{fontenc}    
\usepackage{hyperref}       
\usepackage{url}            
\usepackage{booktabs}       
\usepackage{amsfonts}       
\usepackage{nicefrac}       
\usepackage{microtype}      
\usepackage{graphicx}
\usepackage{doi}
\usepackage{ulem}
\usepackage{graphicx}
\usepackage{geometry}
\usepackage{caption}
\usepackage{subcaption}
\usepackage{amsmath}
\usepackage{amssymb}
\usepackage{cuted}
\usepackage{wasysym}

\usepackage{ulem}
\usepackage[square,numbers]{natbib}
\usepackage{ulem}
\usepackage{graphicx}
\usepackage{geometry}
\usepackage{caption}
\usepackage{subcaption}
\usepackage{cuted}
\usepackage{wasysym}

\usepackage{ulem}

\title{Hierarchical bubble size distributions in coarsening wet liquid foams.}

\author{ 
\hspace{1mm}Nicolo Galvani$^{2,5}$,
\hspace{1mm}Marina Pasquet$^1$,
\hspace{1mm}Arnab Mukherjee$^{2}$,
\hspace{1mm}Alice Requier$^1$,
\hspace{1mm}Sylvie Cohen-Addad$^{2,3}$,\\
\hspace{1mm}\textbf{Olivier Pitois}$^5$,
\hspace{1mm}\textbf{Reinhard Höhler$^{2,3}$},
\hspace{1mm}\textbf{Emmanuelle Rio$^1$},
\hspace{1mm}\textbf{Anniina Salonen$^1$},
\hspace{1mm}\textbf{Douglas J. Durian$^4$},\\
\hspace{1mm}\textbf{Dominique Langevin$^1$}\\
	$^1$ Universit\'e Paris-Saclay, CNRS, Laboratoire de Physique des Solides, 91405, Orsay, France.\\
	$^2$ Sorbonne Université, Institut des NanoSciences de Paris, Paris, France.\\
    $^3$ Université Gustave Eiffel, Laboratoire Navier, Marne-la-Vallée, France. \\
    $^4$ Department of Physics and Astronomy, University of Pennsylvania, Philadelphia, Pennsylvania 19104, USA. \\
    $^5$ Lab Navier, Univ Gustave Eiffel, ENPC, CNRS, 5 Bd Descartes, Champs-sur-Marne,\\
    F-77454Marne-la-Vall\'ee, France. 
}

\renewcommand{\shorttitle}{\textit{Hierarchical bubble size distribution in coarsening wet liquid foams}}

\hypersetup{
pdftitle={Hierarchical bubble size distributions in coarsening wet liquid foams},
pdfauthor={Nicolo Galvani, Marina Pasquet, Arnab Mukherjee, Alice Requier, Sylvie Cohen-Addad, Olivier Pitois, Reinhard Höhler, Emmanuelle Rio, Anniina Salonen, Douglas J. Durian, Dominique Langevin},
pdfkeywords={Foams, Coarsening, Ostwald Ripening},
}

\begin{document}
\maketitle

\begin{abstract}
Coarsening of two-phase systems is crucial for the stability of dense particle packings such as alloys, foams, emulsions or supersaturated solutions. 
Mean field theories predict an asymptotic scaling state with a broad particle size distribution. 
Aqueous foams are good model systems for investigations of coarsening-induced structures, because the continuous liquid as well as the dispersed gas phases are uniform and isotropic. 
We present coarsening experiments on wet foams, with liquid fractions up to their unjamming point and beyond, that are performed under microgravity to avoid gravitational drainage.
As time elapses, a self-similar regime is reached where the normalized bubble size distribution is invariant. 
Unexpectedly, the distribution features an excess of small \textit{roaming} bubbles, mobile within the network of \textit{jammed} larger bubbles. 
These roaming bubbles are reminiscent of rattlers in granular materials (grains not subjected to contact forces).
We identify a critical liquid fraction $\phi^*$, above which the bubble assembly unjams 
and the two bubble populations merge into a single narrow distribution of bubbly liquids. 
Unexpectedly, $\phi^*$ is larger than the random close packing fraction of the foam $\phi_{rcp}$.
This is because, between $\phi_{rcp}$ and $\phi^*$, the large bubbles remain connected due to a weak adhesion between bubbles.
We present models that identify the physical mechanisms explaining our observations. 
We propose a new  comprehensive view of the coarsening phenomenon in wet foams. 
Our results should be applicable to other phase-separating systems and they may also help to control the elaboration of solid foams with hierarchical structures.
\end{abstract}


\section*{Significancy Statement}
Coarsening is a ubiquitous phenomenon in phase separations. 
It is widely observed in alloys, polymers, emulsions, foams and even in biological systems. However, coarsening of materials where the two phases have comparable volume fractions is still poorly understood. To fill this gap, we performed coarsening experiments on aqueous foams in microgravity -- free from gravity-driven destabilization. We discovered that coarsening naturally produces, besides large jammed bubbles, a significant proportion of small \textit{roaming}  bubbles. This hierarchical size distribution is surprising, but could be general and exist in other coarsening systems. Foaming being a generic method to produce solid cellular materials with many applications, making use of these roaming bubbles opens up a new way of designing hierarchical materials.

\section{Introduction}
Liquid phase separation is a common phenomenon in material processing or aging. In the late stages of separation, the  dispersed domains grow, in order to decrease interfacial energy. This often occurs as the dispersed phase diffuses through the continuous phase. This process, known as coarsening or Ostwald ripening, is sometimes referred to as ‘thermodynamic capitalism’ ~\cite{wang2015phase}, where big entities get bigger at the expense of small entities which disappear. 

Coarsening kinetics determines the microstructure of the materials, i.e. the average domain size and size distribution, which is why it has been studied in widely varying contexts. 
The coarsening of alloys has been extensively studied ~\cite{doi:10.1146/annurev.ms.22.080192.001213}, but coarsening also affects crystallization of proteins~\cite{nanev2017recent}, separation by chirality~\cite{noorduin2009ostwald},  synthesis of small particles of controlled microstructure including quantum dots~\cite{asua2018ostwald, huo2010hollow, ross1998coarsening}, stability of foams and emulsions~\cite{Weaire2001,Cantat2013, Langevin2020} as well as other complex liquid-liquid phase separations~\cite{rosowski2020elastic}. Coarsening has recently emerged as an important route to protein compartmentalization within living cells~\cite{Hyman2014}.
\par
The theoretical description of coarsening goes back to Lifshitz and Slyozov~\cite{Lifshitz1961}, and to Wagner~\cite{Wagner1961}, and is usually referred to as ‘LSW theory’. This theory is only valid in the limit  of high volume fraction $\phi$ of the continuous phase. Nonetheless, the LSW theory correctly predicts the coarsening  behavior in numerous systems, namely, that the time-dependent average domain radius increases with time as $t^{1/3}$ and that the normalized size distribution of the domains (PDF) is invariant with time. Many efforts were subsequently  made to account for the behavior at 
smaller $\phi$ by means of theoretical methods and by computer simulations~\cite{baldan2002,yan2022microstructural}.  However, the evolution of the growth regime in $t^{1/3}$ toward the so called 'grain growth regime' in $t^{1/2}$  established for vanishing $\phi$ remains mostly unexplored, both experimentally and theoretically. In view of the large variety of phase separating systems and their existing and potential applications, improving the knowledge on this transition is highly desirable. We have chosen liquid foams as model systems because the domains are fluid and isotropic.
\par
Liquid foams are metastable dispersions of gas bubbles in a liquid matrix ~\cite{Weaire2001, Cantat2013, Langevin2020, ashby1997}. They are not only interesting model systems for coarsening studies, they have numerous practical applications.  Solidifying  the continuous phase of liquid foams yields solid materials which inherit the structure of their precursors. Solid foams are widely used for packaging, insulation or as lightweight construction materials such as foamed cement or metallic foams. Their solid volume fraction is frequently chosen between $20 \%$ and $50\%$, to confer sufficient mechanical strength~\cite{ashby1997}. The solid foam microstructure has an impact on its mechanical properties, for a given density. Hierarchical foam structures were predicted to have an order of magnitude improvement of mechanical strength-to-density ratio with just two levels of hierarchy
~\cite{Lakes1993}. Therefore, such hierarchical structures self-assembled by foam coarsening, as we report here, could be of great interest for applications. 
\par
When the liquid volume fraction $\phi$ is large, the bubbles are spherical and isolated, the dispersions are called ‘bubbly liquids’ rather than foams. Coarsening is expected to lead to a bubble growth proportional to $t^{1/3}$, as in Ostwald ripening.  
When the liquid fraction $\phi$ is decreased below a critical value, $\phi_{\text{rcp}}$, contacts between neighboring bubbles are formed and their shapes progressively evolve from spheres to polyhedra in the limit $\phi \rightarrow 0$~\cite{Cantat2013}. Equilibrium films separating neighboring bubbles have generally thicknesses of a few tens of nanometers. They are connected three by three to channels called \textit{Plateau borders},  themselves connected at \textit{vertices}.  For disordered monodisperse foams, $\phi_{\text{rcp}} \approx 36\%$, and $\phi_{\text{rcp}}$ is expected to decrease slightly as polydispersity increases~\cite{Groot2009}.  
Experiments with 3D foams of small liquid fractions have shown that the average bubble radius grows at long times as $t^{1/2}$ ~\cite{Durian1991, Lambert2010, Isert2013, Chieco2023},
in contrast with the $t^{1/3}$ scaling observed in the case of Ostwald ripening. The value of the exponent is related to the mechanism of gas transfer between bubbles. In dry foams, it  occurs mostly through the thin films between bubbles, whereas in bubbly liquids, gas is transferred through bulk liquid~\cite{Mullins1986}.
Another important feature of coarsening is the shape of the bubble size distribution that is also expected to change with liquid fraction. Several experimental and numerical works~\cite{Feitosa2006,Lambert2007,Magrabi1999,Lambert2010,Thomas2015,Zimnyakov2019} show that the normalized distribution is asymmetric, of the Weibull or lognormal type, in the regime associated with the $t^{1/2}$ growth law, whereas for the $t^{1/3}$ regime, it is more symmetric and narrower~\cite{Lifshitz1961}.
\par
Foams evolve with time not only because of coarsening but also due to gravity drainage~\cite{SAINTJALMES2005, Cantat2013}, and possibly due to rupture of liquid films separating neighboring bubbles, called coalescence. 
Since gravity drainage and coarsening are coupled, studying and modelling coarsening requires gravity drainage to be suppressed. Pioneering foam coarsening experiments were performed with dry horizontal 2D foams (single layers of bubbles) where drainage was not an issue~\cite{stavans1993evolution}. Studies of 3D foams on Earth are generally restricted to small liquid fractions $\phi \ll 0.1$, where drainage is slow enough~\cite{Briceno2017, Durian1991}. \par
To rule out artifacts related to gravity in 3D foams whatever the liquid fraction, we have performed foam coarsening experiments in microgravity, on board the International Space Station (ISS), where drainage is suppressed. Samples with arbitrary liquid volume fractions $\phi$ can thus be studied over long times, up to several days, as required to investigate the Scaling State of foam containing a significant fraction of liquid.

\section{Results and Discussion}

\subsection{Excess of small bubbles}
\label{sec:excess}

\begin{figure*}[ht]
    \centering
    \includegraphics[width=1\linewidth]{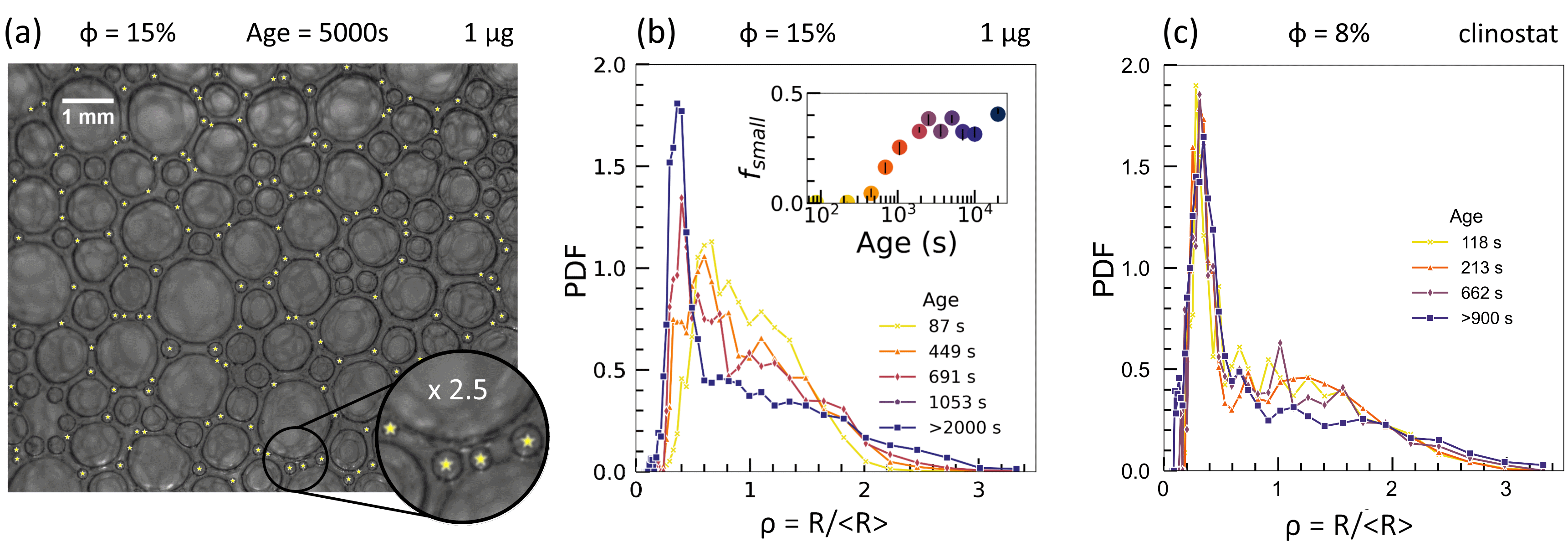}
    \caption{Excess of small bubbles. (a) Image of foam surface ($\phi=15\%$) in the Scaling State regime. Yellow stars have been superimposed on the image to highlight the small bubbles corresponding to the sharp peak in the distribution shown in (b). (b) Probability density function of normalized bubble radius $\rho = R/\langle R \rangle$ at different foam ages as indicated, for a foam with liquid fraction $\phi=15\%$. The curve corresponding to age $ >$ 2000~s represents the Scaling State regime, for which the normalized distribution no longer evolves. Inset: evolution of the proportion of small bubbles as a function of time. The number fraction f$_{\text{small}}$ is obtained by dividing the number of bubbles with radius $R<R_t$ by the total number of bubbles in the sample (see Section~\ref{sec:transition} for details). A change in $R_t$ by $\pm5\%$ induces a variation of f$_{\text{small}}$ smaller than the point size. (c) Probability density function of normalized bubble radius at different ages as indicated, for a sample with liquid fraction $\phi=8\%$ studied on ground. }
    \label{fig:Identification}
\end{figure*}  

We have investigated foam coarsening for liquid fractions between 15\% and 50\% using the instrument described in~\cite{SMD2021}. 
Details can be found in the Materials and Methods section.
From the sample surface observations (a typical image is shown in Fig.~\ref{fig:Identification}a), we measure the bubble sizes using image analysis, and determine the bubble size distributions of the radius normalized by its average $\rho =R/\langle R \rangle$.  The initial size distributions produced by our experimental setup are asymmetric (positive skew) with a maximum at $\rho \approx 0.6$ (see Figure \ref{fig:Identification}b for foam with $15\%$ liquid fraction as an example). The normalized size distributions broaden with time, and a sharp peak builds up progressively for small bubble sizes, \textit{i.e.} $\rho \approx 0.3$,
until a stationary form is reached, indicating a Scaling State. This is shown in Figure \ref{fig:Identification}b for times $t>2000$ s. This evolution is typical of the measurements we have made for foams with liquid fractions within the range  $15 \% \leq \phi < \phi^{\star}$, with $\phi^{\star} \approx 39\%$. 
The small bubbles corresponding to the peak in the distribution are highlighted in Figure \ref{fig:Identification}a. After an increase in the transient regime, they finally represent about $35\%$ of the total bubble population in the scaling state (inset of Figure \ref{fig:Identification}b). We also measured the number of those small bubbles per foam vertex to reach a maximum average value of 1.5, due to space limitation in the vertices. As a consequence, the size distribution becomes invariant in time (statistically self-similar) as observed.

Up to now, such an excess of small bubbles has not been reported in the literature~\cite{Magrabi1999,Feitosa2006,Lambert2010,Zimnyakov2019}. In order to check if  distributions with an excess of small bubbles are also found in drier foams, we have performed coarsening experiments using the same surfactant and a liquid fraction of $8\%$, low enough for gravity effects to be compensated in a ground based experiment by  rotating the cell around a horizontal axis (clinostat). As shown in Figure~\ref{fig:Identification}c we observed a similar excess of small bubbles.
The small bubbles were thus seemingly present, but not detected in previous studies. This is probably because high spatial resolution together with a careful image analysis are needed~\cite{CRAS2023}.  
The only experimental work we have found that indirectly relates to this is that of Feitosa and Durian~\cite{Feitosa2006}, which reports the development of transient bidispersity for initially monodisperse bubbles in a Steady State column, where drainage and coarsening occur simultaneously. In their  simulations of 2D foam coarsening, 
Khakalo \textit{et al}~\cite{Khakalo2018} have observed an excess of small bubbles but the gas transfer through interstitial bulk liquid was not taken into account.
Other peculiar size distributions such as 
``Apollonian” distributions were observed during the decay of beer foam \cite{sauerbrei2006apollonian} and with emulsions \cite{kwok2020apollonian}.
In contrast to what happens in our systems, they arose from coalescence events.

For $\phi > \phi^{\star}$ we have observed a different scenario: the initial bubble size distribution shrinks until a steady state is reached where the size distribution is notably narrow (see Fig. S1 in the SI).  The latter distribution is reminiscent of the theoretical distribution predicted for the Ostwald regime \cite{Lifshitz1961,Wagner1961}. Around $\phi^ *$,  a change in the growth laws for the average bubble size is also observed  for the same foam samples\cite{SMpaper}: 
\begin{equation}
R_{32}^2(t) = R_{32}^2(0)+\Omega_{p} \;t \quad \quad \textrm{for}\quad \phi < \phi^*
\label{eq:bubblegrowth1/2}
\end{equation}
\begin{equation}
R_{32}^3(t) =R_{32}^3(0)+\Omega_{c} \;t\quad \quad \textrm{for}\quad \phi > \phi^*
\label{eq:bubblegrowth1/3}
\end{equation}
 The Sauter mean radius $R_{32} = \langle R^3\rangle/\langle R^2\rangle$ is defined as the ratio of third to second moments of the bubble radius distribution.

\subsection{Transition from jammed bubbles to roaming bubbles} \label{sec:transition}

\begin{figure*}
    \centering
    \includegraphics[width=1\linewidth]{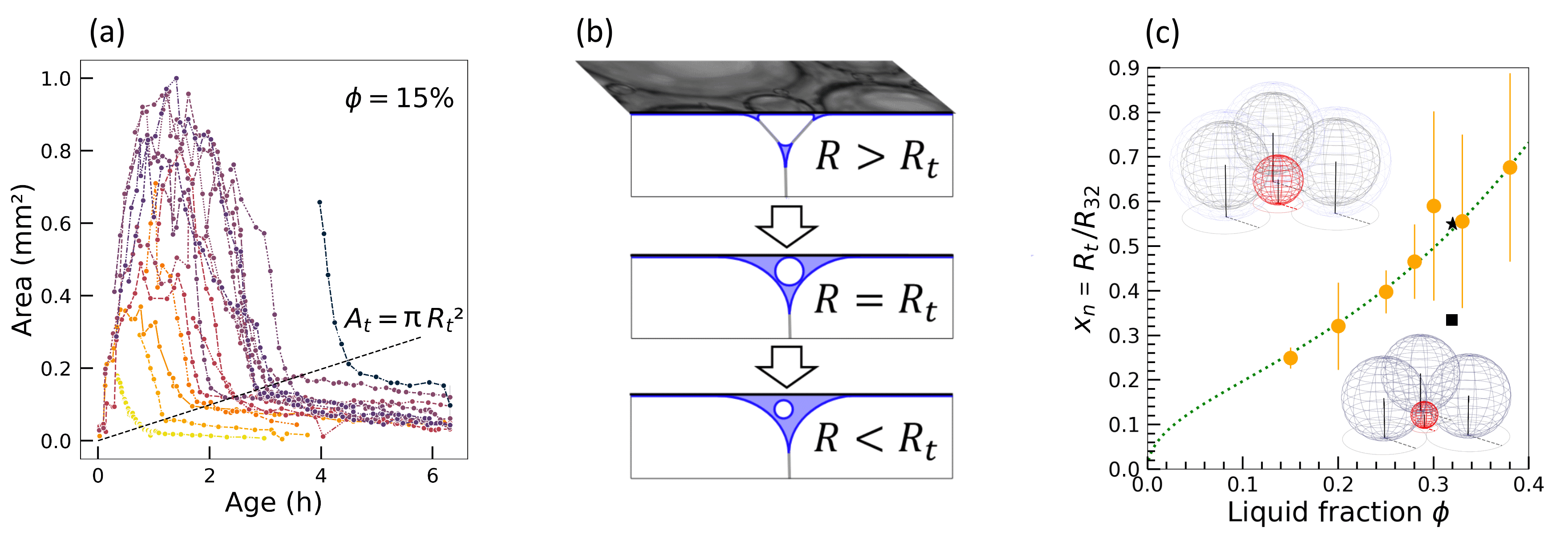}
    \caption{Roaming transition: (a) Evolution of the area of individual bubbles as a function of foam age measured as the time elapsed since the end of the foam sample production, for $\phi=15\%$. The area $A_t=\pi R_t^2$ denotes the bubble area at the wall when its shrinking abruptly slows down (see text). Each label corresponds to a different bubble.
    (b) The transition to the very small shrinking rate was observed to occur when the foam bubble has become so small that it fits inside the interstice between neighboring larger bubbles. The corresponding geometrical transition can therefore be described as follows: When its radius is larger than $R_t$, the small bubble is a foam bubble, in the fact that it shares thin liquid films with its neighbors. In contrast, as its radius reaches values smaller than $R_t$, the bubble loses its contacts with its neighbors: it becomes a roaming bubble and its shrinking rate is strongly decreased.
    (c)  Coefficient $x_n = R_t/R_{32}$ as a function of $\phi$. Filled orange disks: values deduced from the tracking of individual bubbles. Error bars show $\pm 3 SD$, to highlight the observed variability. Black stars/drawings: calculation of $x_n$ from the size of a hard sphere (in red) that can be inserted into the interstice formed by three spheres at the wall, assuming either a compact bubble cage (bottom) or slight loosening (top) of the latter. The dotted line corresponds to equation~\ref{eq:xn} with $\xi = 2.2$.  }
    \label{fig:Roaming}
\end{figure*}

To clarify the origin of the hierarchical bubble population, we have identified bubbles that eventually disappear and tracked evolution of their area. Figure \ref{fig:Roaming}a shows examples of such measurements in a foam with $\phi = 15 \%$. Similar data are shown for other liquid fractions between $\phi = 20 \%$ and $\phi = 33 \%$ in the Supplementary Information (cf.~Figure S2). 
Over time, the individual bubble area can either increase or decrease, depending on the bubble's gas exchanges with its neighbours, but most of the observed bubbles eventually shrink (see Figure \ref{fig:Roaming}a).
The magnitude of the shrinking rate appears to be initially similar to that characterizing the
initial growing rate. 
Then, a transition occurs and the area decreases much more slowly. Actually, the shrinking after this transition can be extremely slow, and we think this is the underlying mechanism explaining why a peak at smaller than average bubbles builds up in the size distribution. Remarkably, the bubble radius at the transition, $R_t$, is such that its area $A_t = \pi R_t^2$ increases linearly with time, which is similar to the evolution of the squared mean radius in the Scaling State (Eq.~\ref{eq:bubblegrowth1/2}). 
Moreover, the transition to the very small shrinking rate appears to occur when the bubble has become so small that it fits inside the interstice between three larger bubbles at the surface, and possibly loses contacts with them as sketched in Fig.~\ref{fig:Roaming}b. (See movies S1-S3 in the SI.) They are free to move throughout the interstice without being pressed against multiple neighbors. Such small bubbles can have different configurations in the interstice, \textit{i.e.} near the center of the interstice or in contact with one bubble or two bubbles, but these configurations do not last for the entire life of the bubbles because their positions are jostled as the jammed bubbles intermittently rearrange due to the coarsening induced dynamics~\cite{CohenAddad2001,Cantat2013}. We call the small bubbles \textit{roaming bubbles}. Note that they are reminiscent of rattlers (grains carrying no force) in granular media~\cite{Agnolin2007}. We conjecture that the bubble size at the transition, $R_t$, should scale as the maximum radius of a sphere that can be trapped in such an interstice at the wall surface. \\

In a coarsening foam that has reached the Scaling State, there is only one independent length scale of the bubble packing structure. Since the bubbles that form the interstices are bigger than the encaged roaming bubbles, we chose to characterize their average size by the Sauter mean radius. With respect to $\langle R \rangle$,  $R_{32}$ indeed represents mainly the average radius of the larger bubbles of the distribution and minimizes the contribution of the small bubbles. 
At a time $t$, the maximum radius of a sphere  trapped in such a vertex can be written, on average:
\begin{equation}
R_t(t, \phi) = x_n(\phi) R_{32} (t)
\label{eq:Rt}
\end{equation}
where $x_n(\phi)$ is a dimensionless geometrical coefficient. We show in Figure~S3 of the SI the plots of $R_t$ versus $R_{32}$ for each liquid fraction. The plots are reasonably described by equation~\ref{eq:Rt}, allowing determination of the average coefficient $x_n$ for each liquid fraction (see Figure \ref{fig:Roaming}c). $x_n(\phi)$ varies from $0.25$ to $0.55$ as $\phi$ varies from $15 \%$ to $38 \%$ respectively. Using those $x_n$ values, the transition radii $R_t$ collapse on a linear master curve when plotted versus $x_n(\phi) R_{32}$ (cf. Figure~S3 of the SI).\\

We have performed a geometrical calculation of the size of the interstice between a plane and three perfect spheres of equal radius $R_{32}$ in contact together and with the plane (see Figure \ref{fig:Roaming}c). This leads to $x_n = 1/3$. This value is smaller than what is measured for liquid fractions corresponding to the bubble random close packing fraction, \textit{i.e.} $\phi_{\text{rcp}} \approx 31 \%$ (see section~\ref{sec:distributions} for more details), beyond which the bubbles are spherical. As the liquid fraction gets close to $\phi_{\text{rcp}}$, the foam osmotic pressure, which pushes neighboring bubbles against each other at contacts, becomes very low, and it can be inferred that the cage formed by the triplets of bubbles of radius $R_{32}$ loosens. Note that such a geometrical loosening effect is general and independent of friction~\cite{delarrard1999}. Therefore, as a correction to the previous calculation, a distance $\epsilon R_{32}$ is added around each sphere (see Figure \ref{fig:Roaming}c). The coefficient now reads: $x_n = \frac{\epsilon (2+\epsilon)+\scriptscriptstyle \frac{4}{3}(1+\epsilon)^2}{2(2+\epsilon)} \approx \frac {1}{3} + \epsilon$, and it increases significantly due to the loosening effect: assuming a moderate loosening $\epsilon \approx 0.$ gives $x_n \approx 0.5$ which is in better agreement with our measurements (see Figure \ref{fig:Roaming}c and movie S2). 
It is reasonable to assume that polydispersity may also impact the size of the interstice. This effect can be estimated by considering two bubbles of size $R_{32}$ and a third one with size $\beta R_{32}$. It can be shown that in such a case, the coefficient reads $x_n \approx \frac {1}{3} + 0.11(\beta-1)$. Therefore, the magnitude of the polydispersity effect is much weaker than the previous one, in addition to the fact that it can work in both directions, depending on the value of $\beta$, which we observed to vary in the range $0.3 < \beta < 1.5$ (see Figure S4 of the SI). However, it is worth noting that a significant fraction of bubbles have a radius larger than $R_{32}$, \textit{i.e.} $1 \leq \beta \leq 1.5$, and that almost half of the nodes are bounded by one such large bubble (see Figure S4 as an example for foam with $15 \%$ liquid). 
These findings suggest that the effect of polydispersity is only slightly positive, and should only slightly increase $x_n$ i.e. the size of the wall interstice. We conclude that the loosening of the bubble packing is the main effect accounting for the measured $x_n$ values.\\

To extend our prediction to any liquid fraction $\phi \leq \phi_{\text{rcp}}$, we turn to~\cite{Louvet2010}, where the radius of passage of a hard sphere through the liquid channels (so-called Plateau borders~\cite{Cantat2013}) was determined as a function of $\phi$ and bubble radius $R$ in a monodisperse foam. Due to uniformity of the capillary pressure through the foam, which sets the radius of curvature of the channels, and thus their cross-section, the bubble radius at the transition $R_t$ should be proportional to this radius of passage. Following the approach proposed in~\cite{Louvet2010} we refer to the effective pore radius introduced by Johnson \textit{et al.}~\cite{Johnson1986}: $\Lambda \approx (8 \overset{\large\sim}{\small{k}} / \overset{\large\sim}{\small{\sigma}})^{1/2}R$, where $\overset{\large\sim}{\small{k}}$ is the dimensionless liquid Darcy's permeability through the foam structure, \textit{i.e.}
$k/R^2$, and $\overset{\large\sim}{\small{\sigma}}$ is the ratio of the electrical conductivity of the foam to that of the foaming liquid. Therefore, the expression sought for $x_n$ is:
\begin{equation}
x_n = \xi (8 \overset{\large\sim}{\small{k}} / \overset{\large\sim}{\small{\sigma}})^\frac{1}{2}
\label{eq:xn}
\end{equation}
where $\xi$ is a geometrical coefficient to be determined. Note that the latter is expected to account for the loosening and polydispersity effects discussed previously. To continue we now need expressions for $\overset{\large\sim}{\small{k}}$ and $\overset{\large\sim}{\small{\sigma}}$. Since $\Lambda$ was initially proposed for solid porous media, the permeability should correspond to foam having rigid interfaces to mimic solid-like boundary conditions. As studied by Rouyer \textit{et al.}~\cite{Rouyer2010}, its expression is given by: $\overset{\large\sim}{\small{k}} = \phi^2 / (312(1-2.15\phi + 1.37\phi^2)^2)$ within the range of liquid fractions $1\% \leq \phi \leq 40\%$. 
For foams and bubbly liquids, Feitosa \textit{et al.}~\cite{Feitosa2005} proposed an approximate analytical expression for $\overset{\large\sim}{\small{\sigma}}$, \textit{i.e.} $\overset{\large\sim}{\small{\sigma}} = 2\phi(1+12\phi)/(6+29\phi -9\phi^2)$. Using these expressions, we set $\xi = 2.2$ in equation~\ref{eq:xn} in order to get a predicted value of $x_n$ close to the measured value 0.53 for $\phi \approx \phi_{\text{rcp}}$ (see Figure~\ref{fig:Roaming}c). Remarkably, the agreement with our experimental data is very good over the whole range of liquid fractions, which reinforces the physical picture that $R_t$ actually corresponds to the size of the interstices formed by the jammed bubbles around the roaming bubbles. Note that in all of the above, nothing is really specific to the fact that we are looking at the wall. In bulk, typical interstices are formed by four bubbles in a tetrahedral assembly. The geometrical calculation for four bubbles in contact gives $\sqrt{\scriptscriptstyle \frac{3}{2}}-1 \approx 0.225$, compared to $1/3$ at the wall. Therefore we can estimate $x_n$ for bulk by using equation \ref{eq:xn} with coefficient $\xi=2.2 \times (0.225/0.333) \approx 1.5$. Provided this value is used, the behavior observed at the wall should be similar to the behavior observed in the bulk of the foam.\\

\subsection{Dissolution rate of the roaming bubbles}

\begin{figure}
    \centering
    \includegraphics[width=0.5\linewidth]{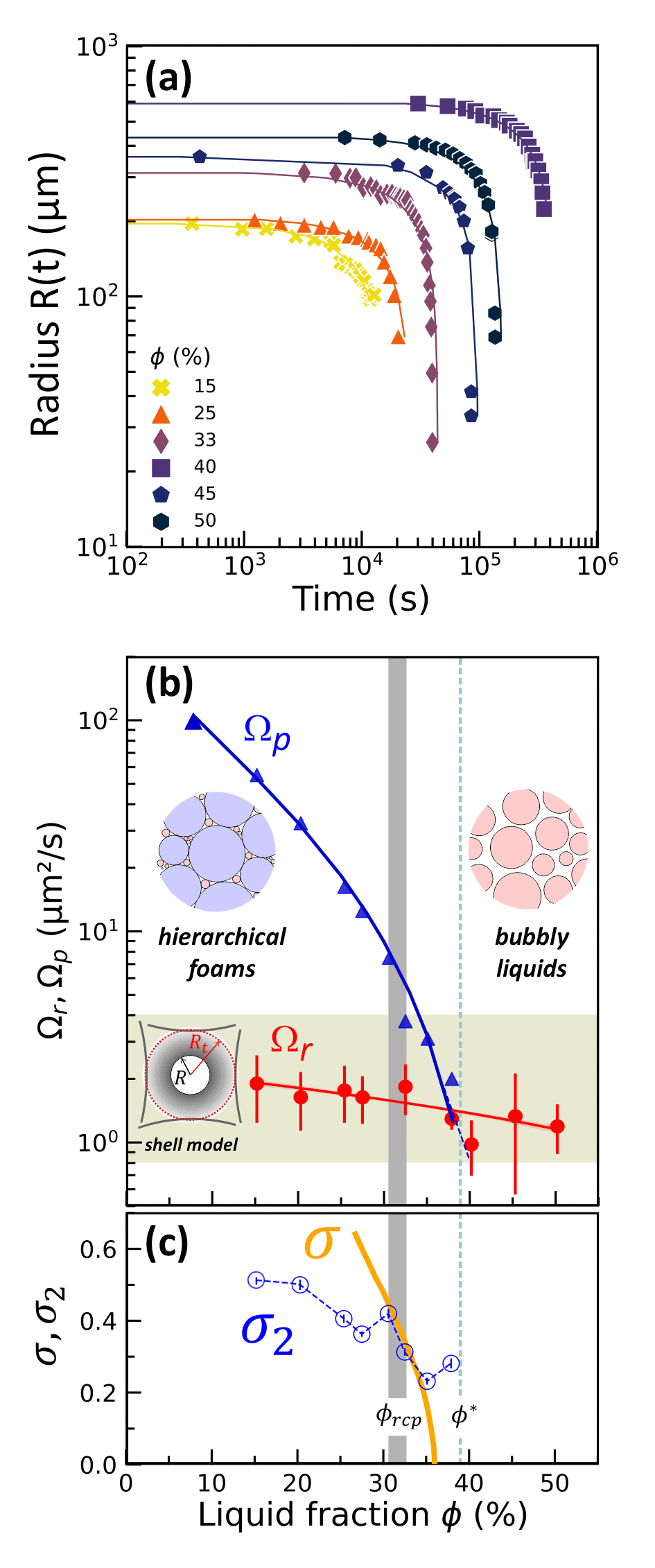}
    \caption{Roaming bubble dissolution: (a) Radius evolution of dissolving roaming bubbles where each curve represents a single bubble. The solid lines correspond to fits of Eq.~\ref{eq:shrinking}. (b) Average shrinking rate of roaming bubbles $\Omega_r$ as a function of liquid fraction compared to the growth rate of average bubble size in the foam $\Omega_{p}$ (Eq.~\ref{eq:bubblegrowth1/2}, data from~\cite{SMpaper}). The lines are guides to the eye. $\Omega_r$ values fall within the range (highlighted in green) predicted by the shell model (Eq.~1 in the SI), schematically illustrated by the inside drawing. Error bars correspond to $\pm 1 SD$. The growth rate $\Omega_{p}$  is strongly dependent on the liquid fraction, at the difference of the dissolution rate $\Omega_r$. (c) Measured shape parameter $\sigma_2$ of the jammed bubbles size distribution (Eq.~6 in the SI) as a function of liquid fraction (blue circles). The (orange) continuous line represents the maximum packing volume fraction predicted for a lognormal distribution of spheres with shape parameter $\sigma$~\cite{Farr2013, Groot2009}. 
    The gray vertical area highlights the range where $\sigma$ and $\sigma_{2}$ coincide, from which we deduce $\phi_{\text{rcp}} \approx 30 - 32 \%$. 
    This also corresponds to the range of liquid fractions where $\Omega_r$ is comparable to $\Omega_{p}$ in b. }
    \label{fig:Dissolution}
\end{figure}

In this section, we focus on the dissolution rate of the roaming bubbles in the range $\phi < \phi^*$. We first consider the data for times longer than those that mark the intersection of the dissolution curve with $A(t) = \pi R_t^2$ (Figures~\ref{fig:Roaming}a and S2 in the SI).
We follow the evolution of the radius of roaming bubbles $R(t)$ for $R(0) \lesssim R_t$. For comparison, we similarly analyze individual bubbles roaming in the bubbly liquids ($\phi > \phi^*$), from the instant they start to continuously shrink.  Several examples of the curves  are presented in Figure~\ref{fig:Dissolution}a. We observe that the following function fits well all the curves\cite{EpsteinPlesset1950,Michelin2018}:
\begin{equation}
R^2(t) = R^2(0) - \Omega_r t
\label{eq:shrinking}
\end{equation}
where the only fitted parameter $\Omega_r$ represents the dissolution rate of the roaming bubble. Such fits were performed for all the liquid fractions and the average values of $\Omega_r$ are presented in Figure~\ref{fig:Dissolution}b. $\Omega_r$ is found to depend only weakly on liquid fraction: $\Omega_r \approx 1-2~\mu$m$^2/s$. We also plot on Figure~\ref{fig:Dissolution}b the growth rate $\Omega_{p}$ that characterizes the coarsening of the foam in the Scaling State (Eq.~\ref{eq:bubblegrowth1/2}).  It appears that $\Omega_{p} \gg \Omega_r$ for $\phi \lesssim \phi_{\text{rcp}}\approx31\%$, and $\Omega_{p} \approx \Omega_r$ for $\phi_{\text{rcp}} < \phi < \phi^*$. 
This comparison reinforces our discussion in section~\ref{sec:excess}: the size of the roaming bubbles, represented on the left side of the distribution, varies more slowly than the average bubble size. As a result, the roaming bubbles accumulate in the interstices formed by the larger bubbles. 
\\

As the dissolution rate $\Omega_r$ plays a crucial role in the accumulation mechanism of the roaming bubbles, we seek here to understand this value. The starting point is the comparison of our data with theory for the  dissolution of isolated bubbles \cite{EpsteinPlesset1950,Michelin2018}, which gives the steady dissolution rate far enough from the final instant of bubble disappearance as: $\Omega_r = -dR^2/dt = 2 D_m V_m \left(c(R) - c_{\infty}\right) = 2 D_m V_m \text{He} P_0 (1-\zeta)$, where the saturation parameter $\zeta = c_{\infty}/\text{He} P_0$ characterizes the gas saturation of the liquid environment, $c(R)$ and $c_{\infty}$ are respectively the gas concentrations in the liquid at the bubble surface and at infinity, $P_0$ is the  gas pressure at infinity,  $\text{He}$ and $D_m$ are respectively the Henry solubility and the diffusion coefficient of the air molecules in the foaming solution. 
$V_m$ is the molar volume of the gas at the pressure $P_0$. 
From the measured $\Omega_r$, we deduce an effective value for the saturation parameter:  $\zeta = 0.973-0.987$, which suggests that the bubbles dissolve faster than if they were isolated, and despite the presence of the large neighbouring bubbles which impose at their interface a gas concentration larger than $\text{He} P_0$. To explain this apparent contradiction, it is important to understand that the gas transfer is controlled by the concentration gradient, and not only by the concentration difference. Due to the short distances involved between the roaming bubble interface and the interfaces of the large neighbouring bubbles, the concentration gradient around the roaming bubble reaches relatively high values compared to the case of the isolated bubble. Therefore, to mimic this situation, we consider the configuration illustrated in the inset of Figure~\ref{fig:Dissolution}b, where a roaming bubble of radius $R$ is centered in a cavity of radius $R_t$, and is surrounded by a liquid shell of thickness $R_t-R$. The \textit{local} concentration at the outside boundary of the shell is estimated as that at the surface of a bubble of average size $R_{32}$. From Fick's first law, we then predict the bubble dissolution rate $\Omega_r$ in that shell environment (See more details in the SI).  For the range of values of $R_{32}$ in the scaling state in our experiments and typical ratio $R/R_t$, we expect $\Omega_r \approx 0.75-4~\mu$m$^2$/s which provides boundaries consistent with the measured values of $\Omega_r$ (cf. Fig.~\ref{fig:Dissolution}b).
 
A drawback of this shell-like model is that the roaming bubble is assumed to remain at the center of the interstice, which is not always the case. Indeed, we often noticed transient apparent contacts between the roaming bubble and either one of the bubbles delimiting the interstice or two larger bubbles forming a corner. These transient contacts can result from adhesive forces. We have indeed observed that under microgravity conditions, persistent aggregates form spontaneously in dilute bubble dispersions. In complementary ground-based experiments, we have observed a contact angle close to $3-4^o$~\cite{SMpaper}. The underlying configuration may be an adhesive contact with the formation of a liquid film that slightly flattens the bubbles or it can be a near-contact with a small separation distance so that the roaming bubble is spherical. Since it was not possible to distinguish between these two types of contact, we estimated the dissolution rate for both cases (See details of the calculation in the SI). 
In the range of average bubble sizes $R_{32}$ of our experiments, assuming a  film thickness effective for the transport of gas of the order of 40-60~nm~\cite{SMpaper}, we found that the expected rates fall within the range of values measured for $\Omega_r$.
This remains broadly true if the bubble is in a corner, where the corresponding dissolution rate is twice larger. Therefore, whatever the configuration considered for the roaming bubble in the interstice, we find values for its dissolution rate that are compatible with our measurements, which gives robustness to the proposed mechanism based on the accumulation of long-lasting roaming bubbles in the foam interstices.\\

\subsection{Bubble size distributions and random close packing fraction in the Scaling State}
\label{sec:distributions}
\begin{figure*}
\centering
\includegraphics[width=0.975\linewidth]{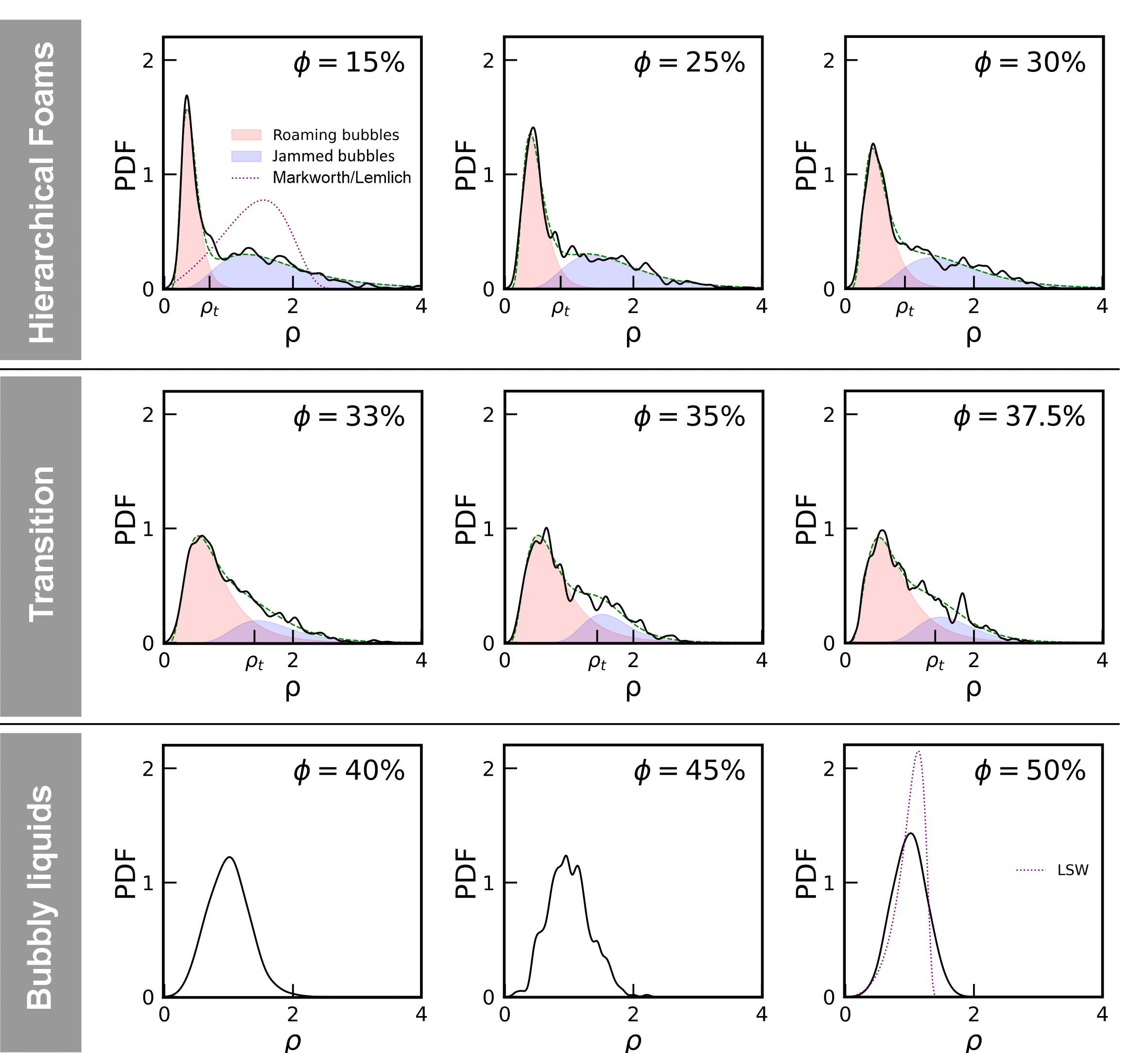}
\caption{Bubble size distributions of normalized radius $\rho = R/\langle R \rangle$ for each liquid fraction as labelled. The data are represented by black continuous lines.  The green dashed lines represent the bi-lognormal PDFs (see Eq.~6 in the SI) fitted  to the data.
The red (resp. blue) shaded area corresponds to the roaming bubble PDF $w \; \mathcal{L}(r; m_1, \sigma_1)$ (resp. to the foam bubble PDF $(1 - w) \; \mathcal{L}(\rho; m_2, \sigma_2)$) with the parameters given in Fig.~S5 of the SI. In the plots for $\phi$ up to $38\%$ the width of the roaming bubble distributions is characterized by $\rho_t$, defined in Eq.~8 in the SI.
For $\phi=15\%$, the  dotted line is the PDF predicted for wet foams by Markworth~\cite{Markwort1985} based on Lemlich's model~\cite{Lemlich1978} for that $\phi$.  As a comparison, for $\phi = 50\%$, the dotted line is the LSW prediction~\cite{baldan2002}  ($\phi = 1$). }
\label{fig:Distributions}
\end{figure*}

Let us analyze now the role of liquid fraction on the distribution shape. Details on the analysis are given in the SI. 
Figure~\ref{fig:Distributions} shows the normalized bubble size distributions observed in the Scaling State for each sample liquid fraction.

The PDF for $\phi=15\%$ is the same as that of Figure 1 in the Scaling State.
It exhibits a prominent narrow peak, that we identified to the roaming bubble population in Section~\ref{sec:excess}, followed by a broad peak for the foam bubble population.  These features qualitatively persist up to $\phi<38\%$ but the narrow peak progressively shifts towards larger $\rho$ while its height decreases. For $\phi \geq 40\%$, PDFs exhibit a single peak, which is consistent with the fact that all bubbles should be roaming bubbles. PDFs become narrower as $\phi$ increases and their peak height increases. This qualitative change is also captured by the abrupt variation of statistical quantities like polydispersity and standard deviation (cf. Fig.~S5 of the SI). None of the existing theories predict such distributions~\cite{baldan2002}. These findings indicate a cross-over between qualitatively different PDFs occurring for a liquid fraction $\phi^{\star} \approx 39\%$. This transition coincides with the observed change of growth laws Eq.~\ref{eq:bubblegrowth1/2} and Eq.~\ref{eq:bubblegrowth1/3}  and it is attributed to the onset of the formation of a foam gel due to weak attraction between bubbles as evidenced by finite contact angle at films junctions~\cite{SMpaper}.\\

The expected jamming liquid fraction  for randomly close-packed monodisperse hard spheres is $\phi_{\text{rcp}} = 36\%$. However, polydispersity will
reduce this value  since smaller bubbles can fit into the interstices
between larger ones. 
This effect has been predicted by  numerical simulations of polydisperse close packings of spherical particles with lognormal PDF, as a function of the shape parameter $\sigma$~\cite{Farr2013, Groot2009}. 
In our foams, the close packing concerns the population of jammed bubbles, which are connected to each other \textit{via} films. Therefore, we compare the measured shape parameter of the foam bubble distribution $\sigma_2$ to the predicted ones (cf. Fig.~\ref{fig:Dissolution}c). We find them to coincide within the range $\phi=30\%$ and $\phi=32\%$: we 
expect the close packing fraction $\phi_{\text{rcp}}$ of our foams to lay inside the range between these 2 values.\par

To provide an independent result of the close packing fraction of frictionless spheres with the polydispersity observed in our samples in the Scaling State, we have performed molecular dynamics simulations. Since here we are only interested in the geometrical sphere packing problem at the jamming point where the confinement pressure and interaction forces drop to zero with increasing $\phi$, we expect the nature of the interaction law used in the simulations to have only a minor impact. 
Using Hertzian interactions, in the framework of the molecular dynamics code LAMMPS (see Materials and Methods), we obtained $\phi_{\text{rcp}}=$31.0$\pm0.5\%$, in remarkable agreement with our analysis based on the work of Farr and Groot~\cite{Groot2009}. Note that strictly speaking, our simulations only provide an upper bound for the optimal random close packing fraction of such polydisperse spheres, which may be obtained by more sophisticated simulation procedures described in the literature~\cite{Kansal2002}. However, in the context of our experiments, the truly relevant packing fraction is the one of a coarsening foam. In this case we expect a local packing which is not exactly the most compact possible one. A jammed foam regularly undergoes rearrangements, helping it to settle into  new minimal energy configurations. This implies that in between rearrangements, the packing is not always optimally close-packed. Simulations of this where we also replace the Hertzian interaction by the more realistic Morse Witten law~\cite{morse1993droplet,HOHLER2019} are the subject of ongoing work.   

\par

\subsection{Potential consequences on foam properties}

\label{sec:potential}

The roaming bubbles represent a significant proportion of the total number of bubbles. As a result, exclusion of the roaming bubbles from the determination of the average radius leads to an up to 3-fold overestimation of coarsening rates, depending on the liquid fraction~\cite{SMpaper}. 

In terms of volume fraction with respect to the liquid volume, the roaming bubbles represent up to ten percent depending on $\phi$. It can therefore be expected that their impact is important for certain properties.
Roaming bubbles can also modify foam drainage, where they slow down the flow of the liquid. A study with solid spheres, located in the nodes of the liquid network of the foam, showed that such an amount of particles in the liquid could reduce the permeability of the foam by $40\%$~\cite{Rouyer2014}. 
This shows the bias of systematically ignoring their presence.
Let us mention that to date, this effect has never been taken into account in permeability modelling.

The presence of the roaming bubbles can be highly detrimental for applications of foams where the microstructure is an important parameter, all the more so if they migrate and accumulate in large proportions in certain places.
Moreover, for a number of these applications, the interstitial liquid is a complex fluid, possibly with yield stress properties that will prevent gravity from evacuating the roaming bubbles: one expects to find relatively high volume fractions of roaming bubbles in such systems.

By examining some recent papers on foamed construction materials, we recognised traces of the presence of such roaming bubbles. For example in~\cite{Feneuil2019}, very small pores can be seen in the bulk nodes of cement foam solidified after coarsening, as revealed by the micro-tomography image.
There is every reason to believe that literature is full of such examples.\par
On the other hand, one can make use of these roaming bubbles. Note that the stakes are high in terms of producing solid foam structures with hierarchical porosity, including both macro- and micro-scaled pores. Such structures have recently been produced by 3D printing~\cite{Chen2018} and they were found to present enhanced energy absorption properties and enhanced mechanical resistance to cyclic loading. 
\par

\section*{Conclusions}
Studies of foam samples where the liquid fraction remains constant over periods of several days, without any confounding effects of gravitational drainage, reveal that, as demonstrated earlier for dry foams, wet foams evolve towards a Scaling State. In this state, the bubble size distributions show a well-defined peak towards  smaller than average  sizes, i.e. an excess of bubbles for sizes close to $0.3\langle R \rangle$. This feature is not predicted by existing theories. During coarsening, as shrinking bubbles become smaller than the size of interstices between the larger bubbles, they can move independently from the jammed bubble network to become \textit{roaming} bubbles. Surprisingly, although we have been able to reproduce this effect on Earth, no previous experimental study mentions the presence of these small bubbles, except for a study of draining foams~\cite{Feitosa2006} but where roaming bubbles have not been identified as such.This suggests that, although their study has been underestimated, hierarchical bubble size distributions can build up on Earth if drainage is not too fast compared to dissolution and coarsening.
\par

The dissolution rate of these roaming bubbles is approximately constant, whatever the liquid fraction of the samples. The dissolution rate is consistent with calculations based on the gas transfer through the liquid shell that surrounds the roaming bubble, or through the ``contact`` between one roaming bubble and larger jammed bubbles surrounding them. The key point in the accumulation of the small bubbles in the interstices formed by the larger bubbles, is the fact that the rate of disappearance of these bubbles is much smaller than the average growth rate of the jammed bubbles. This behaviour is observed for foams with liquid fractions smaller than the random close packing fraction $\phi_{\text{rcp}}$. 
For $\phi$ between $\phi_{\text{rcp}}$ and $\phi^{\star}$, where the bubble assembly  approaches the regime of \textit{bubbly liquids}, the rate of dissolution of the roaming bubbles reaches progressively the growth rate of the jammed bubbles, which suppresses the accumulation mechanism. As a consequence, the peak initially observed for liquid fractions $\phi < \phi_{\text{rcp}}$  shifts towards $\langle R \rangle$ and a distribution almost centered on $\langle R \rangle$, characteristic of bubbly liquids, is eventually observed. For $\phi$ above $\phi*$, none of the bubbles are confined.

In closing, we have shown the existence of naturally developed hierarchical bubble size distributions in coarsening  foams. 
We present a comprehensive view of coarsening of wet foams, completely different from expectations, with a persistent co-existence of jammed bubbles with small roaming bubbles, and the existence of a range between $\phi_{\text{rcp}}$  and $\phi*$ where foam bubbles are still jammed although not close-packed. These findings challenge our current understanding of foam coarsening and has potential implications in the design and performance of foamy materials. This view should not be restricted to foams but also be applicable to other two-phase systems driven by interfacial effects, such as emulsions, alloys and binary fluid/polymer mixtures. It should be mentioned that recent studies of alloys  with small volume fraction of the continuous phase suggest that the Ostwald ripening regime persists when $\phi$ is smaller than $\phi_{\text{rcp}}$ ~\cite{yan2022microstructural}. The difference between coarsening of foams and alloys remains to be clarified.

\section*{Materials and Methods}
The foams were made with aqueous solutions of a ionic surfactant, tetradecyl-trimethyl-ammonium bromide (TTAB), with purity $\ge 99$~\% and used as received from Sigma-Aldrich. It was dissolved at 5~g/L in ultrapure water (resistivity 18.2~M$\Omega \cdot$cm). This concentration is 4 times larger than the critical micellar concentration and large enough to prevent coalescence. The surface tension of the TTAB solution measured at room temperature is: $\gamma= 37.1$~mN/m. The Henry solubility coefficient of the air molecules in the foaming solution is~\cite{SMpaper} $He=7.4~10^{-6}$~mol~m$^{-3}$~Pa$^{-1}$ ~\cite{SMpaper} and their diffusion coefficient in the foaming solution is~\cite{SMpaper} $D_m = 2.0~10^{-9}$~m$^2$s$^{-1}$.\par
The majority of the experiments were performed on board the International Space Station using the experiment container described in~\cite{SMD2021}. In this environment, the residual gravity acceleration fluctuations are reported to be on the order of or less than a µg, for frequencies below 0.01 Hz \cite{NASAacceleration}.
Each foam cell was filled on Earth with a given volume of foaming solution (measured by weight at controlled temperature) and air, then hermetically sealed. The liquid volume fraction $\phi$ contained in each cell was deduced from the liquid volume and the total cell volume. 
After the completion of the experiments, the cells were send back to Earth and we checked that their weight had  varied by less than 1\%. All the experiments were repeated three times and found reproducible, even a few months apart.
\par
In addition, we made a Ground experiment with a foaming liquid of composition identical to that of the ISS experiments. The foams were produced with the double syringe method~\cite{CRAS2023} filled with air and a volume of the foaming solution in order to set the liquid fraction to 7.8\%$\pm 0.2 \%$. Note that the initial bubble size distribution with this foam production is close to that of the Scaling State.  The sample was placed in a cylindrical cell (diameter 30 mm, thickness 12.8 mm) with transparent flat faces. The cell was kept  with its symmetry axis aligned in the horizontal direction and rotated about this axis with a speed of rotation equal to 15 rpm. 
\par
Foam age is counted from the instant when the foaming process stops.
Bubbles at the surface of the sample are recorded using a video camera. Every image (such as the one shown in Fig.~\ref{fig:Identification}a) was analysed as described in~\cite{CRAS2023}. We checked that the radial profile of liquid fraction remained constant throughout the measurement duration, indicating that the effect of gravity drainage was indeed counteracted and that the rotation did not induce radial drainage either.The bubble area $A$ deduced was from the area inside the contour of the bubbles measured using the ellipses method~\cite{CRAS2023}. Finally, the bubble radius is calculated as $R = \sqrt{A/\pi}$. In the ISS experiments, simultaneously to the video recording, the intensity of light transmitted through the sample was recorded, which provided the average bubble size in the bulk of the sample as explained in~\cite{CRAS2023}. Our results showed that the evolution of the average bubble radius measured either at the surface or in the bulk are similar.
\par
We also performed numerical simulations to evaluate the random close packing liquid fraction of the bubbles. In the framework of the molecular dynamics code LAMMPS \cite{LAMMPS2002}, a cubic simulation box was filled by spheres with repulsive, Hertzian interactions with radii randomly chosen from a distribution corresponding to the one we observe experimentally for $\phi=33\%$ in the Scaling State (see Figure~\ref{fig:Distributions}). The number of spheres was of the order of 2000, similar to our foam coarsening experiments at the largest investigated foam ages. To fill the simulation cell, we started with an initial cell volume so large that the sphere dispersion was highly diluted. Using the pressostat provided by LAMMPS, we then shrunk the cubic cell and compacted these structures until a very small osmotic pressure appeared. We then turned off the pressostat and equilibrated the sample for imposed simulation box volumes, varied by small steps around the previous value. 
 The close packing fraction was estimated by plotting confinement pressure versus packing fraction, and by detecting the $\phi$ value where zero pressure is reached within numerical accuracy. We did this for 5 different initial random seeds, and found $\phi_{\text{rcp}} = $31.0$ \pm 0.5 \%$.
The way you compact a packing has a large impact on the final close packing fraction in frictional granular materials, and to a lesser extent also in frictionless systems. To investigate this effect, we applied simulated gravity to dilute sphere dispersions as an alternative to the initial pressostat procedure. Kinetic energy was dissipated by introducing viscous friction in the contact law. This procedure mimics foams that form when a bubbly liquid is subjected to buoyancy, as it is common on earth. Once equilibrium was reached, we switched off gravity, and simulated pressure versus packing fraction as previously. The final values of $\phi_{\text{rcp}}$ are within statistical errors the same as the those obtained with the pressostat.

\section*{Acknowledgments}
We acknowledge funding by ESA and CNES (via the projects “Hydrodynamics of Wet Foams”) focused on the Soft Matter Dynamics instrument and the space mission Foam-C, as well as NASA via grant number 80NSSC21K0898. Marina Pasquet, Nicolo Galvani and Alice Requier benefited from CNES and ESA PhD grants. The authors are grateful to the BUSOC team for their invaluable help during the ISS experiments. We also want to warmly thank Marco Braibanti and Sébastien Vincent-Bonnieu from ESA, Christophe Delaroche from CNES and Olaf Schoele-Schulz from Airbus for their continuing support.

\bibliography{Biblio}  

\begin{thebibliography}{61}
\providecommand{\natexlab}[1]{#1}
\providecommand{\url}[1]{\texttt{#1}}
\expandafter\ifx\csname urlstyle\endcsname\relax
  \providecommand{\doi}[1]{doi: #1}\else
  \providecommand{\doi}{doi: \begingroup \urlstyle{rm}\Url}\fi

\bibitem[Agnolin and Roux(2007)]{Agnolin2007}
I.~Agnolin and J.-N. Roux.
\newblock Internal states of model isotropic granular packings. i. assembling
  process, geometry, and contact networks.
\newblock \emph{Physical Review E}, 76\penalty0 (6):\penalty0 061302, 2007.
\newblock \doi{10.1103/PhysRevE.76.061302}.
\newblock URL \url{https://link.aps.org/doi/10.1103/PhysRevE.76.061302}.
\newblock PRE.

\bibitem[Asua(2018)]{asua2018ostwald}
J.~M. Asua.
\newblock Ostwald ripening of reactive costabilizers in miniemulsion
  polymerization.
\newblock \emph{European Polymer Journal}, 106:\penalty0 30--41, 2018.

\bibitem[Baldan(2002)]{baldan2002}
A.~Baldan.
\newblock Review progress in ostwald ripening theories and their applications
  to the $\gamma$'-precipitates in nickel-base superalloys part ii nickel-base
  superalloys.
\newblock \emph{Journal of materials science}, 37\penalty0 (12):\penalty0
  2379--2405, 2002.

\bibitem[Born et~al.(2021)Born, Braibanti, Cristofolini, Cohen-Addad, Durian,
  Egelhaaf, Escobedo-Sánchez, Höhler, Karapantsios, Langevin, Liggieri,
  Pasquet, Rio, Salonen, Schröter, Sperl, Sütterlin, and
  Zuccolotto-Bernez]{SMD2021}
P.~Born, M.~Braibanti, L.~Cristofolini, S.~Cohen-Addad, D.~J. Durian, S.~U.
  Egelhaaf, M.~A. Escobedo-Sánchez, R.~Höhler, T.~D. Karapantsios,
  D.~Langevin, L.~Liggieri, M.~Pasquet, E.~Rio, A.~Salonen, M.~Schröter,
  M.~Sperl, R.~Sütterlin, and A.~B. Zuccolotto-Bernez.
\newblock Soft matter dynamics: A versatile microgravity platform to study
  dynamics in soft matter.
\newblock \emph{Review of Scientific Instruments}, 92\penalty0 (12):\penalty0
  124503, 2021.
\newblock \doi{10.1063/5.0062946}.
\newblock URL \url{https://doi.org/10.1063/5.0062946}.

\bibitem[Briceño-Ahumada and Langevin(2017)]{Briceno2017}
Z.~Briceño-Ahumada and D.~Langevin.
\newblock On the influence of surfactant on the coarsening of aqueous foams.
\newblock \emph{Advances in Colloid and Interface Science}, 244:\penalty0
  124--131, 2017.
\newblock ISSN 0001-8686.
\newblock \doi{https://doi.org/10.1016/j.cis.2015.11.005}.
\newblock URL
  \url{https://www.sciencedirect.com/science/article/pii/S0001868615001967}.
\newblock Special Issue in Honor of the 90th Birthday of Prof. Eli Ruckenstein.

\bibitem[Cantat et~al.(2013)Cantat, Cohen-Addad, Elias, Graner, H{\"o}hler,
  Pitois, Rouyer, and Saint-Jalmes]{Cantat2013}
I.~Cantat, S.~Cohen-Addad, F.~Elias, F.~Graner, R.~H{\"o}hler, O.~Pitois,
  F.~Rouyer, and A.~Saint-Jalmes.
\newblock \emph{Foams: structure and dynamics}.
\newblock OUP Oxford, 2013.

\bibitem[Chen et~al.(2018)Chen, Cao, and Advincula]{Chen2018}
Q.~Chen, P.-F. Cao, and R.~C. Advincula.
\newblock Mechanically robust, ultraelastic hierarchical foam with tunable
  properties via 3d printing.
\newblock \emph{Advanced Functional Materials}, 28\penalty0 (21):\penalty0
  1800631, 2018.
\newblock \doi{https://doi.org/10.1002/adfm.201800631}.
\newblock URL
  \url{https://onlinelibrary.wiley.com/doi/abs/10.1002/adfm.201800631}.

\bibitem[Chieco and Durian(2023)]{Chieco2023}
A.~T. Chieco and D.~J. Durian.
\newblock A simply solvable model capturing the approach to statistical
  self-similarity for the diffusive coarsening of bubbles, droplets, and
  grains.
\newblock \emph{arXiv:2303.09612}, 2023.

\bibitem[Cohen-Addad and Höhler(2001)]{CohenAddad2001}
S.~Cohen-Addad and R.~Höhler.
\newblock Bubble dynamics relaxation in aqueous foam probed by multispeckle
  diffusing-wave spectroscopy.
\newblock \emph{Physical Review Letters}, 86\penalty0 (20):\penalty0
  4700--4703, 2001.
\newblock URL \url{http://prl.aps.org/abstract/PRL/v86/i20/p4700_1}.
\newblock TY - JOUR.

\bibitem[De~Larrard(1999)]{delarrard1999}
F.~De~Larrard.
\newblock \emph{Concrete Mixture Proportioning: A Scientific Approach}.
\newblock CRC Press, 1999.
\newblock ISBN 9780429179099.
\newblock \doi{10.1201/9781482272055}.

\bibitem[Durian et~al.(1991)Durian, Weitz, and Pine]{Durian1991}
D.~Durian, D.~Weitz, and D.~Pine.
\newblock Multiple light-scattering probes of foam structure and dynamics.
\newblock \emph{Science}, 252\penalty0 (5006):\penalty0 686--688, 1991.

\bibitem[Epstein and Plesset(1950)]{EpsteinPlesset1950}
P.~S. Epstein and M.~S. Plesset.
\newblock On the stability of gas bubbles in liquid-gas solutions.
\newblock \emph{The Journal of Chemical Physcis}, 18\penalty0 (11):\penalty0
  1505--1509, 1950.
\newblock \doi{10.1063/1.1747520}.

\bibitem[Farr(2013)]{Farr2013}
R.~S. Farr.
\newblock Random close packing fractions of lognormal distributions of hard
  spheres.
\newblock \emph{Powder Technology}, 245:\penalty0 28--34, 2013.
\newblock ISSN 0032-5910.
\newblock \doi{10.1016/j.powtec.2013.04.009}.
\newblock URL \url{https://dx.doi.org/10.1016/j.powtec.2013.04.009}.

\bibitem[Farr and Groot(2009)]{Groot2009}
R.~S. Farr and R.~D. Groot.
\newblock Close packing density of polydisperse hard spheres.
\newblock \emph{The Journal of Chemical Physics}, 131\penalty0 (24):\penalty0
  244104, 2009.
\newblock \doi{10.1063/1.3276799}.

\bibitem[Feitosa et~al.(2005)Feitosa, Marze, Saint-Jalmes, and
  Durian]{Feitosa2005}
K.~Feitosa, S.~Marze, A.~Saint-Jalmes, and D.~J. Durian.
\newblock Electrical conductivity of dispersions: from dry foams to dilute
  suspensions.
\newblock \emph{Journal of Physics: Condensed Matter}, 17\penalty0
  (41):\penalty0 6301, 2005.
\newblock \doi{10.1088/0953-8984/17/41/001}.

\bibitem[Feitosa et~al.(2006)Feitosa, Halt, Kamien, and Durian]{Feitosa2006}
K.~Feitosa, O.~L. Halt, R.~D. Kamien, and D.~J. Durian.
\newblock Bubble kinetics in a steady-state column of aqueous foam.
\newblock \emph{Europhysics Letters ({EPL})}, 76\penalty0 (4):\penalty0
  683--689, nov 2006.
\newblock \doi{10.1209/epl/i2006-10304-5}.
\newblock URL \url{https://doi.org/10.1209/epl/i2006-10304-5}.

\bibitem[Feneuil et~al.(2019)Feneuil, Aimedieu, Scheel, Perrin, Roussel, and
  Pitois]{Feneuil2019}
B.~Feneuil, P.~Aimedieu, M.~Scheel, J.~Perrin, N.~Roussel, and O.~Pitois.
\newblock Stability criterion for fresh cement foams.
\newblock \emph{Cement and Concrete Research}, 125:\penalty0 105865, 2019.
\newblock \doi{10.1016/j.cemconres.2019.105865}.
\newblock URL \url{https://doi.org/10.1016/j.cemconres.2019.105865}.

\bibitem[Gibson and Ashby(1997)]{ashby1997}
L.~J. Gibson and M.~F. Ashby.
\newblock \emph{The mechanics of foams: basic results}, page 175–234.
\newblock Cambridge Solid State Science Series. Cambridge University Press, 2
  edition, 1997.
\newblock \doi{10.1017/CBO9781139878326.007}.

\bibitem[Huisman and Mysels(1969)]{Mysels1969}
F.~Huisman and K.~J. Mysels.
\newblock The contact angle and the depth of the free-energy minimum in thin
  liquid films. their measurement and interpretation.
\newblock \emph{The Journal of Physical Chemistry}, 73:\penalty0 489--497,
  1969.
\newblock \doi{10.1021/J100723A004}.

\bibitem[Huo et~al.(2010)Huo, Wang, Irran, Yu, Gao, Fan, Li, Wang, Ding, Amin,
  et~al.]{huo2010hollow}
J.~Huo, L.~Wang, E.~Irran, H.~Yu, J.~Gao, D.~Fan, B.~Li, J.~Wang, W.~Ding,
  A.~M. Amin, et~al.
\newblock Hollow ferrocenyl coordination polymer microspheres with micropores
  in shells prepared by ostwald ripening.
\newblock \emph{Angewandte Chemie}, 122\penalty0 (48):\penalty0 9423--9427,
  2010.

\bibitem[Hyman et~al.(2014)Hyman, Weber, and Jülicher]{Hyman2014}
A.~A. Hyman, C.~A. Weber, and F.~Jülicher.
\newblock Liquid-liquid phase separation in biology.
\newblock \emph{Annual Review of Cell and Developmental Biology}, 30\penalty0
  (1):\penalty0 39–58, 2014.
\newblock ISSN 1081-0706.
\newblock \doi{10.1146/annurev-cellbio-100913-013325}.
\newblock URL \url{https://dx.doi.org/10.1146/annurev-cellbio-100913-013325}.

\bibitem[Höhler and Weaire(2019)]{HOHLER2019}
R.~Höhler and D.~Weaire.
\newblock Can liquid foams and emulsions be modeled as packings of soft elastic
  particles?
\newblock \emph{Advances in Colloid and Interface Science}, 263:\penalty0
  19--37, 2019.
\newblock ISSN 0001-8686.
\newblock \doi{https://doi.org/10.1016/j.cis.2018.11.002}.
\newblock URL
  \url{https://www.sciencedirect.com/science/article/pii/S0001868618302744}.

\bibitem[Isert et~al.(2013)Isert, Maret, and Aegerter]{Isert2013}
N.~Isert, G.~Maret, and C.~M. Aegerter.
\newblock Coarsening dynamics of three-dimensional levitated foams: {From} wet
  to dry.
\newblock \emph{The European Physical Journal E}, 36\penalty0 (10):\penalty0
  116, Oct. 2013.
\newblock ISSN 1292-8941, 1292-895X.
\newblock \doi{10.1140/epje/i2013-13116-x}.
\newblock URL \url{http://link.springer.com/10.1140/epje/i2013-13116-x}.

\bibitem[Johnson et~al.(1986)Johnson, Koplik, and Schwartz]{Johnson1986}
D.~L. Johnson, J.~Koplik, and L.~M. Schwartz.
\newblock New pore-size parameter characterizing transport in porous media.
\newblock \emph{Physical Review Letters}, 57\penalty0 (20):\penalty0
  2564--2567, 1986.
\newblock \doi{10.1103/PhysRevLett.57.2564}.
\newblock URL \url{https://doi.org/10.1103/PhysRevLett.57.2564}.

\bibitem[Kansal et~al.(2002)Kansal, Torquato, and Stillinger]{Kansal2002}
A.~R. Kansal, S.~Torquato, and F.~H. Stillinger.
\newblock Computer generation of dense polydisperse sphere packings.
\newblock \emph{The Journal of Chemical Physics}, 117\penalty0 (18):\penalty0
  8212--8218, 2002.
\newblock \doi{10.1063/1.1511510}.
\newblock URL \url{https://doi.org/10.1063/1.1511510}.

\bibitem[Khakalo et~al.(2018)Khakalo, Baumgarten, Tighe, and
  Puisto]{Khakalo2018}
K.~Khakalo, K.~Baumgarten, B.~Tighe, and A.~Puisto.
\newblock Coarsening and mechanics in the bubble model for wet foams.
\newblock \emph{Physical Review E}, 98:\penalty0 012607, 2018.
\newblock \doi{10.1103/PhysRevE.98.012607}.
\newblock URL \url{https://dx.doi.org/10.1103/PhysRevE.98.012607}.

\bibitem[Kwok et~al.(2020)Kwok, Botet, Sharpnack, and
  Cabane]{kwok2020apollonian}
S.~Kwok, R.~Botet, L.~Sharpnack, and B.~Cabane.
\newblock Apollonian packing in polydisperse emulsions.
\newblock \emph{Soft Matter}, 16\penalty0 (10):\penalty0 2426--2430, 2020.

\bibitem[Lakes(1993)]{Lakes1993}
R.~Lakes.
\newblock Materials with structural hierarchy.
\newblock \emph{nature}, 361\penalty0 (6412):\penalty0 511,515, 1993.
\newblock \doi{10.1038/361511a0}.

\bibitem[Lambert et~al.(2007)Lambert, Cantat, Delannay, Mokso, Cloetens,
  Glazier, and Graner]{Lambert2007}
J.~Lambert, I.~Cantat, R.~Delannay, R.~Mokso, P.~Cloetens, J.~A. Glazier, and
  F.~m.~c. Graner.
\newblock Experimental growth law for bubbles in a moderately ``wet'' 3d liquid
  foam.
\newblock \emph{Phys. Rev. Lett.}, 99:\penalty0 058304, Aug 2007.
\newblock \doi{10.1103/PhysRevLett.99.058304}.
\newblock URL \url{https://link.aps.org/doi/10.1103/PhysRevLett.99.058304}.

\bibitem[Lambert et~al.(2010)Lambert, Mokso, Cantat, Cloetens, Glazier, Graner,
  and Delannay]{Lambert2010}
J.~Lambert, R.~Mokso, I.~Cantat, P.~Cloetens, J.~A. Glazier, F.~m.~c. Graner,
  and R.~Delannay.
\newblock Coarsening foams robustly reach a self-similar growth regime.
\newblock \emph{Phys. Rev. Lett.}, 104:\penalty0 248304, Jun 2010.
\newblock \doi{10.1103/PhysRevLett.104.248304}.
\newblock URL \url{https://link.aps.org/doi/10.1103/PhysRevLett.104.248304}.

\bibitem[Langevin(2020)]{Langevin2020}
D.~Langevin.
\newblock \emph{Emulsions, Microemulsions and Foams}.
\newblock Springer International Publishing, 2020.
\newblock ISBN 978-3-030-55680-8.
\newblock \doi{10.1007/978-3-030-55681-5}.

\bibitem[Lemlich(1978)]{Lemlich1978}
R.~Lemlich.
\newblock Prediction of changes in bubble size distribution due to interbubble
  gas diffusion in foam.
\newblock \emph{Industrial \& Engineering Chemistry Fundamentals}, 17\penalty0
  (2):\penalty0 89--93, 1978.
\newblock \doi{10.1021/i160066a003}.
\newblock URL \url{https://doi.org/10.1021/i160066a003}.

\bibitem[Lifshitz and Slyozov(1961)]{Lifshitz1961}
I.~Lifshitz and V.~Slyozov.
\newblock The kinetics of precipitation from supersaturated solid solutions.
\newblock \emph{Journal of Physics and Chemistry of Solids}, 19\penalty0
  (1):\penalty0 35--50, 1961.
\newblock ISSN 0022-3697.
\newblock \doi{https://doi.org/10.1016/0022-3697(61)90054-3}.
\newblock URL
  \url{https://www.sciencedirect.com/science/article/pii/0022369761900543}.

\bibitem[Louvet et~al.(2010)Louvet, Hohler, and Pitois]{Louvet2010}
N.~Louvet, R.~Hohler, and O.~Pitois.
\newblock Capture of particles in soft porous media.
\newblock \emph{PHYSICAL REVIEW E}, 82:\penalty0 041405, 2010.
\newblock \doi{https://doi.org/10.1103/PhysRevE.82.041405}.

\bibitem[Magrabi et~al.(1999)Magrabi, Dlugogorski, and Jameson]{Magrabi1999}
S.~Magrabi, B.~Dlugogorski, and G.~Jameson.
\newblock Bubble size distribution and coarsening of aqueous foams.
\newblock \emph{Chemical Engineering Science}, 54:\penalty0 4007--4022, 1999.
\newblock \doi{10.1016/S0009-2509(99)00098-6}.

\bibitem[Markworth(1985)]{Markwort1985}
A.~J. Markworth.
\newblock Comments on foam stability, ostwald ripening, and grain growth.
\newblock \emph{Journal of Colloid and Interface Science}, 107:\penalty0
  569--571, 1985.

\bibitem[Michelin et~al.(2018)Michelin, Guérin, and Lauga]{Michelin2018}
S.~Michelin, E.~Guérin, and E.~Lauga.
\newblock Collective dissolution of microbubbles.
\newblock \emph{PHYSICAL REVIEW FLUIDS}, 043601:\penalty0 043601, 2018.
\newblock \doi{10.1103/PhysRevFluids.3.043601}.

\bibitem[Morse and Witten(1993)]{morse1993droplet}
D.~Morse and T.~Witten.
\newblock Droplet elasticity in weakly compressed emulsions.
\newblock \emph{Europhysics Letters}, 22\penalty0 (7):\penalty0 549, 1993.

\bibitem[Mullins(1986)]{Mullins1986}
W.~W. Mullins.
\newblock The statistical self‐similarity hypothesis in grain growth and
  particle coarsening.
\newblock \emph{Journal of Applied Physics}, 59\penalty0 (4):\penalty0
  1341--1349, 1986.
\newblock \doi{10.1063/1.336528}.
\newblock URL \url{https://doi.org/10.1063/1.336528}.

\bibitem[Nanev(2017)]{nanev2017recent}
C.~N. Nanev.
\newblock Recent experimental and theoretical studies on protein
  crystallization.
\newblock \emph{Crystal Research and Technology}, 52\penalty0 (1):\penalty0
  1600210, 2017.

\bibitem[NASA(2015)]{NASAacceleration}
NASA.
\newblock Acceleration environment, 2015.
\newblock URL
  \url{https://www.nasa.gov/sites/default/files/atoms/files/acceleration-environment-iss-mini-book_detail-508c.pdf}.

\bibitem[Noorduin et~al.(2009)Noorduin, Vlieg, Kellogg, and
  Kaptein]{noorduin2009ostwald}
W.~L. Noorduin, E.~Vlieg, R.~M. Kellogg, and B.~Kaptein.
\newblock From ostwald ripening to single chirality.
\newblock \emph{Angewandte Chemie International Edition}, 48\penalty0
  (51):\penalty0 9600--9606, 2009.

\bibitem[of~Mathematics(2010)]{KStest}
E.~of~Mathematics.
\newblock Kolmogorov–smirnov test, 2010.
\newblock URL
  \url{http://encyclopediaofmath.org/index.php?title=Kolmogorov\%E2\%80\%93Smirnov_test\&oldid=22660}.

\bibitem[Pasquet et~al.(2023{\natexlab{a}})Pasquet, Galvani, Pitois,
  Cohen-Addad, H{\"o}hler, Chieco, Dillavou, Hanlan, Durian, Rio,
  et~al.]{CRAS2023}
M.~Pasquet, N.~Galvani, O.~Pitois, S.~Cohen-Addad, R.~H{\"o}hler, A.~T. Chieco,
  S.~Dillavou, J.~M. Hanlan, D.~J. Durian, E.~Rio, et~al.
\newblock Aqueous foams in microgravity, measuring bubble sizes.
\newblock \emph{Comptes Rendus. M{\'e}canique}, 351\penalty0 (S2):\penalty0
  1--23, 2023{\natexlab{a}}.

\bibitem[Pasquet et~al.(2023{\natexlab{b}})Pasquet, Galvani, Requier, Pitois,
  Höhler, Rio, Salonen, and Langevin]{SMpaper}
M.~Pasquet, N.~Galvani, S.~Requier, A.and Cohen-Addad, O.~Pitois, R.~Höhler,
  E.~Rio, A.~Salonen, and D.~Langevin.
\newblock Coarsening transitions of wet liquid foams under microgravity
  conditions.
\newblock \emph{arXiv:2304.11206}, 2023{\natexlab{b}}.

\bibitem[Rosowski et~al.(2020)Rosowski, Sai, Vidal-Henriquez, Zwicker, Style,
  and Dufresne]{rosowski2020elastic}
K.~A. Rosowski, T.~Sai, E.~Vidal-Henriquez, D.~Zwicker, R.~W. Style, and E.~R.
  Dufresne.
\newblock Elastic ripening and inhibition of liquid--liquid phase separation.
\newblock \emph{Nature physics}, 16\penalty0 (4):\penalty0 422--425, 2020.

\bibitem[Ross et~al.(1998)Ross, Tersoff, and Tromp]{ross1998coarsening}
F.~Ross, J.~Tersoff, and R.~Tromp.
\newblock Coarsening of self-assembled ge quantum dots on si (001).
\newblock \emph{Physical Review Letters}, 80\penalty0 (5):\penalty0 984, 1998.

\bibitem[Rouyer et~al.(2010)Rouyer, Pitois, Lorenceau, , and
  Louvet]{Rouyer2010}
F.~Rouyer, O.~Pitois, E.~Lorenceau, , and N.~Louvet.
\newblock Permeability of a bubble assembly: From the very dry to the wet
  limit.
\newblock \emph{PHYSICS OF FLUIDS}, 22\penalty0 (4):\penalty0 043302, 2010.
\newblock \doi{10.1063/1.3364038}.
\newblock URL \url{https://doi.org/10.1063/1.3364038}.

\bibitem[Rouyer et~al.(2014)Rouyer, Haffner, Louvet, Khidas, and
  Pitois]{Rouyer2014}
F.~Rouyer, B.~Haffner, N.~Louvet, Y.~Khidas, and O.~Pitois.
\newblock Foam clogging.
\newblock \emph{Soft Matter}, 10:\penalty0 6990--6998, 2014.
\newblock \doi{https://doi.org/10.1039/c4sm00496e}.

\bibitem[Saint-Jalmes et~al.(2005)Saint-Jalmes, Peugeot, Ferraz, and
  Langevin]{SAINTJALMES2005}
A.~Saint-Jalmes, M.-L. Peugeot, H.~Ferraz, and D.~Langevin.
\newblock Differences between protein and surfactant foams: Microscopic
  properties, stability and coarsening.
\newblock \emph{Colloids and Surfaces A: Physicochemical and Engineering
  Aspects}, 263\penalty0 (1):\penalty0 219--225, 2005.
\newblock ISSN 0927-7757.
\newblock \doi{https://doi.org/10.1016/j.colsurfa.2005.02.002}.
\newblock URL
  \url{https://www.sciencedirect.com/science/article/pii/S0927775705001226}.
\newblock A collection of papers presented at the 5th European Conference on
  Foams, Emulsions, and Applications, EUFOAM 2004, University of
  Marne-la-Vallee, Champs sur Marne (France), 5-8 July, 2004.

\bibitem[Sauerbrei et~al.(2006)Sauerbrei, Ha{\ss}, and
  Plath]{sauerbrei2006apollonian}
S.~Sauerbrei, E.~Ha{\ss}, and P.~Plath.
\newblock The apollonian decay of beer foam bubble size distribution and the
  lattices of young diagrams and their correlated mixing functions.
\newblock \emph{Discrete Dynamics in Nature and Society}, 2006, 2006.

\bibitem[Schimming and Durian(2017)]{Schimming2017}
C.~D. Schimming and D.~J. Durian.
\newblock Border-crossing model for the diffusive coarsening of two-dimensional
  and quasi-two-dimensional wet foams.
\newblock \emph{Phys. Rev. E}, 96:\penalty0 032805, Sep 2017.
\newblock \doi{10.1103/PhysRevE.96.032805}.
\newblock URL \url{https://link.aps.org/doi/10.1103/PhysRevE.96.032805}.

\bibitem[Stavans(1993)]{stavans1993evolution}
J.~Stavans.
\newblock The evolution of cellular structures.
\newblock \emph{Reports on progress in physics}, 56\penalty0 (6):\penalty0 733,
  1993.

\bibitem[Thomas et~al.(2015)Thomas, Belmonte, Graner, Glazier, and {de
  Almeida}]{Thomas2015}
G.~L. Thomas, J.~M. Belmonte, F.~Graner, J.~A. Glazier, and R.~M. {de Almeida}.
\newblock 3d simulations of wet foam coarsening evidence a self similar growth
  regime.
\newblock \emph{Colloids and Surfaces A: Physicochemical and Engineering
  Aspects}, 473:\penalty0 109--114, 2015.
\newblock ISSN 0927-7757.
\newblock \doi{https://doi.org/10.1016/j.colsurfa.2015.02.015}.
\newblock URL
  \url{https://www.sciencedirect.com/science/article/pii/S0927775715001387}.
\newblock A Collection of Papers Presented at the 10th Eufoam Conference,
  Thessaloniki, Greece,7-10 July, 2014.

\bibitem[Thompson et~al.(2022)Thompson, Aktulga, Berger, Bolintineanu, Brown,
  Crozier, in~'t Veld, Kohlmeyer, Moore, Nguyen, Shan, Stevens, Tranchida,
  Trott, and Plimpton]{LAMMPS2002}
A.~P. Thompson, H.~M. Aktulga, R.~Berger, D.~S. Bolintineanu, W.~M. Brown,
  P.~S. Crozier, P.~J. in~'t Veld, A.~Kohlmeyer, S.~G. Moore, T.~D. Nguyen,
  R.~Shan, M.~J. Stevens, J.~Tranchida, C.~Trott, and S.~J. Plimpton.
\newblock {LAMMPS} - a flexible simulation tool for particle-based materials
  modeling at the atomic, meso, and continuum scales.
\newblock \emph{Comp. Phys. Comm.}, 271:\penalty0 108171, 2022.
\newblock \doi{10.1016/j.cpc.2021.108171}.

\bibitem[Voorhees(1992)]{doi:10.1146/annurev.ms.22.080192.001213}
P.~W. Voorhees.
\newblock Ostwald ripening of two-phase mixtures.
\newblock \emph{Annual Review of Materials Science}, 22\penalty0 (1):\penalty0
  197--215, 1992.
\newblock \doi{10.1146/annurev.ms.22.080192.001213}.

\bibitem[Wagner(1961)]{Wagner1961}
C.~Wagner.
\newblock Theorie der alterung von niederschl\"agen durch uml\"osen.
\newblock \emph{Zeitchrift f\"ur Elektrochemie}, 65:\penalty0 581--591, 1961.
\newblock \doi{10.1002/bbpc.19610650704}.

\bibitem[Wang and Glicksman(2015)]{wang2015phase}
K.~Wang and M.~Glicksman.
\newblock Phase coarsening in thin films.
\newblock \emph{JOM}, 67:\penalty0 1905--1912, 2015.

\bibitem[Weaire and Hutzler(2001)]{Weaire2001}
D.~L. Weaire and S.~Hutzler.
\newblock \emph{The physics of foams}.
\newblock Oxford University Press, 2001.

\bibitem[Yan et~al.(2022)Yan, Wang, and Glicksman]{yan2022microstructural}
H.~Yan, K.~Wang, and M.~Glicksman.
\newblock Microstructural coarsening in dense binary systems.
\newblock \emph{Acta Materialia}, 233:\penalty0 117964, 2022.

\bibitem[Zimnyakov et~al.(2019)Zimnyakov, Zemlyanukhin, Yuvchenko, Bochkarev,
  Slavnetskov, Gavrilov, and Tumachev]{Zimnyakov2019}
D.~Zimnyakov, A.~Zemlyanukhin, S.~Yuvchenko, A.~Bochkarev, I.~Slavnetskov,
  S.~Gavrilov, and D.~Tumachev.
\newblock Self-similarity of bubble size distributions in the aging metastable
  foams.
\newblock \emph{Physica D}, 398:\penalty0 171--182, 2019.
\newblock \doi{10.1016/j.physd.2019.03.008}.

\end{thebibliography}

\clearpage
\begin{center}
\textbf{\large Supplementary Information}
\end{center}
\setcounter{equation}{0}
\setcounter{figure}{0}
\setcounter{table}{0}
\setcounter{page}{1}
\makeatletter
\renewcommand{\theequation}{S\arabic{equation}}
\renewcommand{\thefigure}{S\arabic{figure}}
\renewcommand{\bibnumfmt}[1]{[S#1]}
\renewcommand{\citenumfont}[1]{S#1}

\section*{Dissolution rate models}
We consider the configuration illustrated in the inset of Figure~3b, where a roaming bubble of radius $R$ is centered in a cavity of radius $R_t$ and  surrounded by a liquid shell of thickness $R_t-R$.
Denoting $r$ the radial distance from the bubble center, the  gas concentrations at $r = R$ and $r = R_t$  are imposed by the bubble Laplace pressures, i.e. $c(R)= He \; (P_0 + 2\gamma/R)$ and $c(R_t)=He \; (P_0 + 2\gamma/R_{32})$, where the \textit{local}  concentration at the outside boundary of the shell is estimated as that at the surface of a bubble of average size $R_{32}$. With these boundary conditions, the dissolved gas concentration profile is only a function of the radial coordinate $r$. It is a solution of the steady diffusion equation and is given by $\frac{c(r)-c(R)}{c(R_t)-c(R)}=\frac{R_t}{R_t-R} (1-R/r)$. From Fick's first law, the molar rate of gas transfer outwards from the roaming bubble is equal to $-4\pi R^2 D_m (dc/dr)_{r=R}$, and the resulting dissolution rate is:
\begin{equation}
\Omega_r = \frac{4 \gamma D_m V_m He}{R} \frac{(\frac{1}{R}-\frac{1}{R_{32}})}{(\frac{1}{R}-\frac{1}{R_t})}
\label{eq:shell}
\end{equation}
 
Values provided by equation~\ref{eq:shell} depend on estimates of the different radii that have been introduced. We choose $R_{32}$ within the range of values corresponding to our experiment in the Scaling State, i.e. $300~\mu m \lesssim R_{32} \lesssim 500~\mu m$ ~\cite{SMpaper}, and we limit the ratio $R/R_t$ within the range $0.2-0.8$ (for $R/R_t \lesssim 0.2$ the size of roaming bubbles is measured with less precision, and for $R/R_t \gtrsim 0.8$ it is difficult to be sure that the tracked bubble is still a roaming bubble). Thus, we get  values for $\Omega_r$ within the range $0.75-4~\mu$m$^2$/s, as represented in Figure~3b.
\\

As explained in the Main text, we noticed transient apparent contacts between the roaming bubble and either one of the bubbles delimiting the interstice or two larger bubbles forming a corner. The underlying configuration may be a real contact, i.e. with the formation of a liquid film that slightly flattens the bubbles, or it can be a near-contact with a small separation distance, in which case the roaming bubble remains spherical. Since it is not possible to distinguish between these two types of contact, we estimate the dissolution rate for both cases.
First we consider the case of the near-contact. We refer to the work of Schimming~$\&$ Durian~\cite{Schimming2017}, who considered the so-called \textit{kissing bubbles} configuration, where the distance $h$ between the two spherical bubbles is such that $h/R \ll 1$. We assume two bubbles of radii $R \approx R_t/2 \approx R_{32}/6$ and $R_{32}$ (radius of surrounding jammed bubbles). The dissolution rate of the small bubble then writes:
\begin{equation}
\Omega_r \approx 2\gamma D_m  V_m He \left(\frac{1}{R}-\frac{1}{R_{32}} \right) \times \ln\left(0.8+0.6\frac{R_{32}}{3\;h}\right) \;
 \approx \frac{10\gamma D_m  V_m He}{R_{32}} \ln\left(0.8+0.6\frac{R_{32}}{3\;h}\right)
\label{eq:Schimming}
\end{equation}
Because of the logarithm values provided by equation~\ref{eq:Schimming} are weakly dependent on $h$. A typical value for $\Omega_r$ is $2~\mu$m$^2$/s.  \\

Now we consider the case of a small bubble of radius $R$ and a big bubble of radius  $R_{32}$ sharing a film, of thickness $h$ and area $A$, which meets the free surface of the bubbles with a contact angle $\theta$~\cite{Mysels1969} (cf. Fig.~\ref{fig:adhesive_bubbles}).
The dissolution rate of the small bubble is set by the Laplace pressure difference $\Delta P$ between both bubbles, which drives the gas transfer through the contact film and sets its curvature. It writes:
\begin{equation}
\Omega_r \approx \frac{ D_m  V_m He}{2\pi\;R\;h} \Delta P \; A
\label{eq:adhesion}
\end{equation}
The film is a spherical cap surface with its radius of curvature  $R^{\star}$ (cf. Fig.~\ref{fig:adhesive_bubbles}) given by: 
\begin{equation}
\Delta P=\frac{4\gamma}{R^{\star}} = 2 \gamma \left(\frac{1}{R}-\frac{1}{R_{32}} \right)
\label{eq:film_curvature}
\end{equation}

The surface area $A = 2 \pi R^{\star 2} \left( 1 - cos(\psi/2) \right)$. Taking the same range of bubble radii as above $R \approx R_t/2 \approx R_{32}/6$ and for the small contact angles considered here,  geometrical considerations shows that at leading order $\psi/2 \approx 5\theta/7$, thus  $A \approx \frac{\pi}{2} R^{\star 2}\theta^2$. Then the dissolution rate(Eq.~\ref{eq:adhesion2}) writes:
\begin{equation}
\Omega_r \approx \frac{12\gamma D_m  V_m He}{ 5\;h} \;\theta^2
\label{eq:adhesion2}
\end{equation}

Values provided by Eq.~\ref{eq:adhesion2} depend on the film thickness and contact angle. For our foams, we have $\theta \approx 4^{\circ}$~\cite{SMpaper}. We remark that an effective film thickness in the range $h\approx40-60$~nm gives a dissolution rate close to the rate calculated for the \textit{kissing bubbles} configuration, i.e. $\approx 2.6-4~\mu$m$^2$/s. Note that the effective thickness accounts for both the aqueous core film thickness and the effective length related to gas transfer resistance from the two surfactant monolayers. Therefore, $h\approx40-60$~nm may well correspond to either a Newton Black Film (NBF) with significant monolayer effect, or to a Common Black Film (CBF) with limited monolayer effect. It seems difficult to go further in this quantitative analysis but it is in good agreement with the effective thickness obtained from the foam coarsening rate with the same solution~\cite{SMpaper}.\\

To estimate the dissolution rate when the bubble is in a corner, we can, as a first approximation, multiply the previous values by a factor of two. For the \textit{kissing bubbles} configuration we get a dissolution rate equal to the upper range of values provided by the shell-model, i.e. $4~\mu$m$^2$/s. Values for roaming bubbles sharing two liquid films with neighboring larger bubbles exceed a little bit that range, i.e. $5-8~\mu$m$^2$/s. As we did not measure such high values, it suggests that such a configuration, if it really exists, is rather rare.\\

Therefore, whatever the configuration considered for the roaming bubble in the interstices, we find values for its dissolution rate that are compatible with our measurements, which gives robustness to the proposed mechanism based on the accumulation of long-lasting roaming bubbles in the foam interstices.

\section*{Analysis of bubble size distributions}

 We first transformed the measured discrete histograms into continuous PDFs using a Gaussian Kernel estimation (with bandwidth equal to 2.5 the image pixel size $\approx 14~\mu$m).
We then described the two bubble populations (either roaming bubble or jammed bubble) in the domain $\phi< \phi^{\star}$ with a bimodal lognormal PDF defined as:
\begin{equation}
    \mathcal{F} = w \cdot \mathcal{L}(\rho; m_1, \sigma_1) + (1 - w) \cdot \mathcal{L}(\rho; m_2, \sigma_2)
    \label{eq:2log}
\end{equation}
where $w$ is the proportion of roaming bubbles in the foam, and $\mathcal{L}$ is a lognormal distribution parameterized as:
\begin{equation}\label{eq:log}
    \mathcal{L}(\rho; m, \sigma) =  \frac{1}{\rho \sigma \sqrt{2 \pi}} \exp{\left[ - \frac{\log^2{(\rho/m)}}{2\sigma^2}\right]}.
\end{equation}
where the shape parameter $\sigma$ is the log-scale standard deviation and $m$ is the linear-scale median. We fitted the bimodal function to the measured PDFs, and observed that the parameter $m_2 = 1.66\pm 0.03 (SEM)$ is almost independent of the liquid fraction. In the following, we fix  $m_2 = 1.66$ and fit the other parameters. Their variations with liquid fraction are given in Fig.~S5. 
The results of the Kolmogorov-Smirnov statistical test~\cite{KStest} applied to the bimodal lognormal fits provide quantitative evidence for fit quality: no fit could be rejected according to the $5\%$ rule, and many are above 40$\%$ (with the exceptions of $p(\phi=15\%)=21\%$ and $p(\phi=25\%)=14\%$).\\

Finally, since $R_t$ should represent the maximum size of the roaming bubbles, we conjecture that it should be correlated to the width of the roaming bubble PDF. To test this, we measure the width at the foot of the PDF  $\mathcal{L}(\rho; m_1, \sigma_1)$ by the normalized radius $\rho_t$ estimated such that number of roaming bubbles with $\rho >\rho_t$ equals the number of jammed bubbles with $\rho < \rho_t$:
\begin{equation}
    (1-w)\int_0^{\rho_t} \mathcal{L}(\rho; m_2, \sigma_2) = w\int_{\rho_t}^\infty \mathcal{L}(\rho; m_1, \sigma_1). 
    \label{eq:rhot}
\end{equation}
We have estimated $\rho_t (\phi)$ up to $\phi=\phi^*$, and in Figure~S6 we compare it to $x_n(\phi)$. Since we employed $R_{32}$ as the reference radius in the estimation of $x_n$, we must rescale $\rho_t$ by $R_{32}$ for a comparison with $x_n$. In the range of liquid fractions up to $\phi=\phi_{\text{rcp}}$, we find $\rho_t \; \langle R \rangle /R_{32} \approx x_n$, which is consistent. 





\section*{Additional Figures}
The additional figures support discussions in the Main text. We did not include text there, the captions being self-explanatory.

\begin{figure}
    \centering
    \includegraphics[width=0.6\linewidth]
    {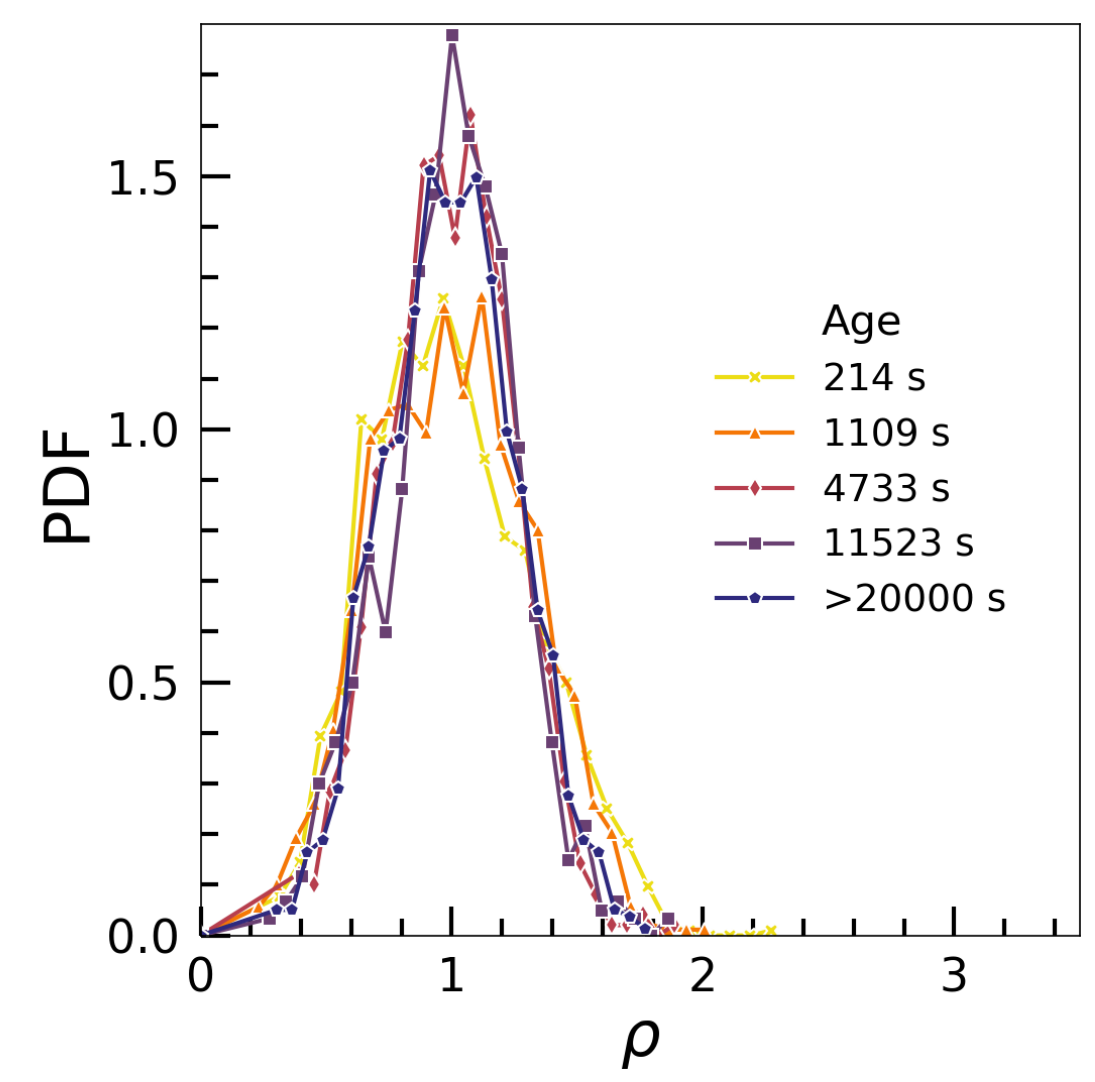}
    \caption{Probability density function of the normalized bubble radius $\rho = R/\langle R \rangle$ at different foam ages as indicated, for a foam with liquid fraction $\phi=50\%$. The curve corresponding to age $ >$ 20000~s represents the Scaling State regime (observed up to 300000s - end of the experiment), for which the normalized distribution no longer evolves. This distribution is an example of a concentrated bubbly liquid, with a single peak and a narrow distribution with bubble sizes strictly smaller than $\rho = 2$.}
    \label{fig:Evolution of phi_50}
\end{figure}

\begin{figure}[ht]
    \centering
    \includegraphics[width=0.8\linewidth]{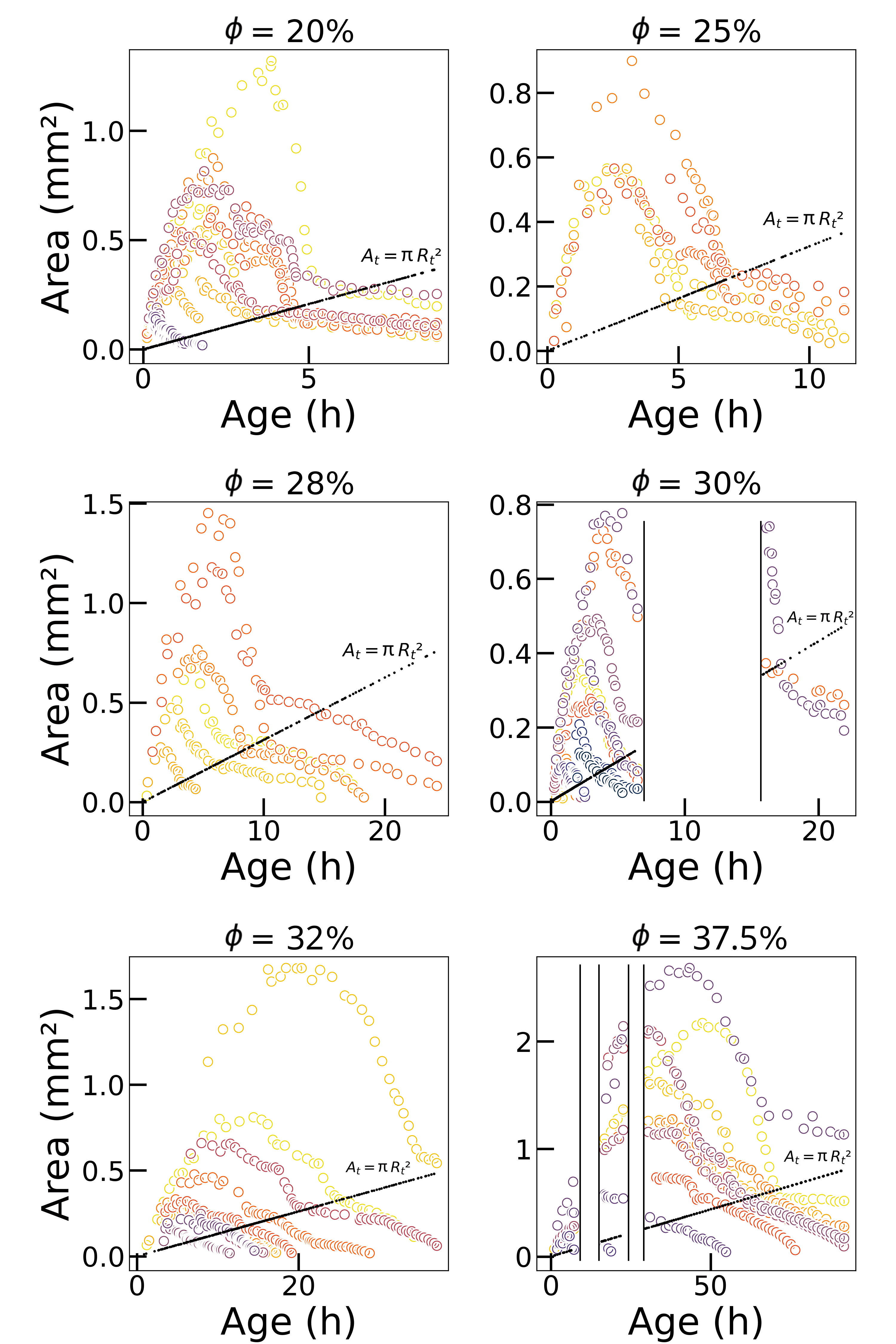}
  \caption{Evolution of the area of individual bubbles as a function of foam age versus the time elapsed since the end of the foam sample production, for a series of liquid fractions. The area $A_t=\pi R_t^2$ denotes the bubble area when its shrinking abruptly slows down (see text). Each color corresponds to the evolution of a different bubble. For samples with $\phi = 30\%$  or $38\%$, data were acquired in parallel with other samples. As a consequence, there are some blanks in the image in between  the vertical lines.
  }
  \label{fig:SI_evolution_small} 
\end{figure}

\begin{figure}
    \centering
    \includegraphics[width=0.499\linewidth]{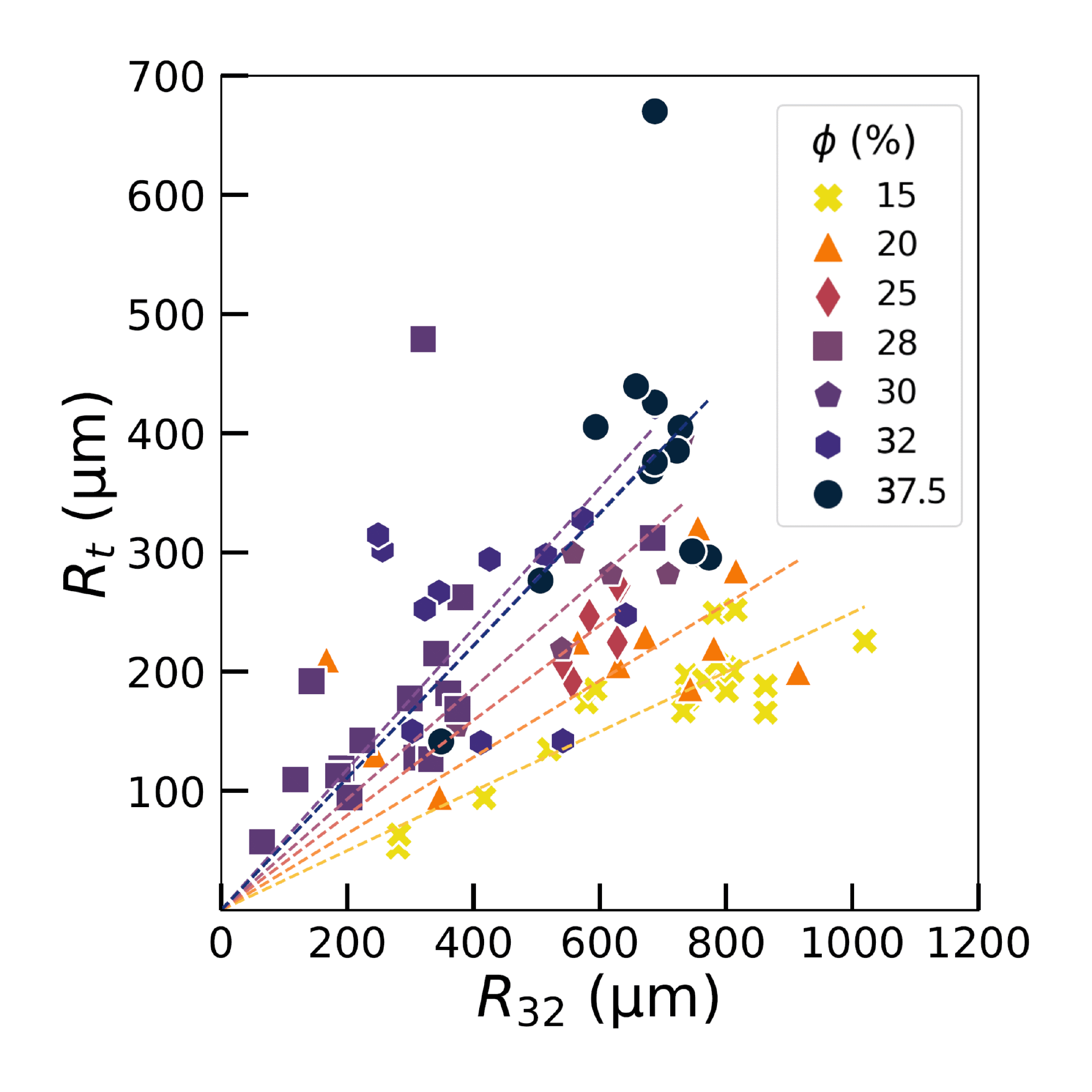}
    \includegraphics[width=0.4955\linewidth]{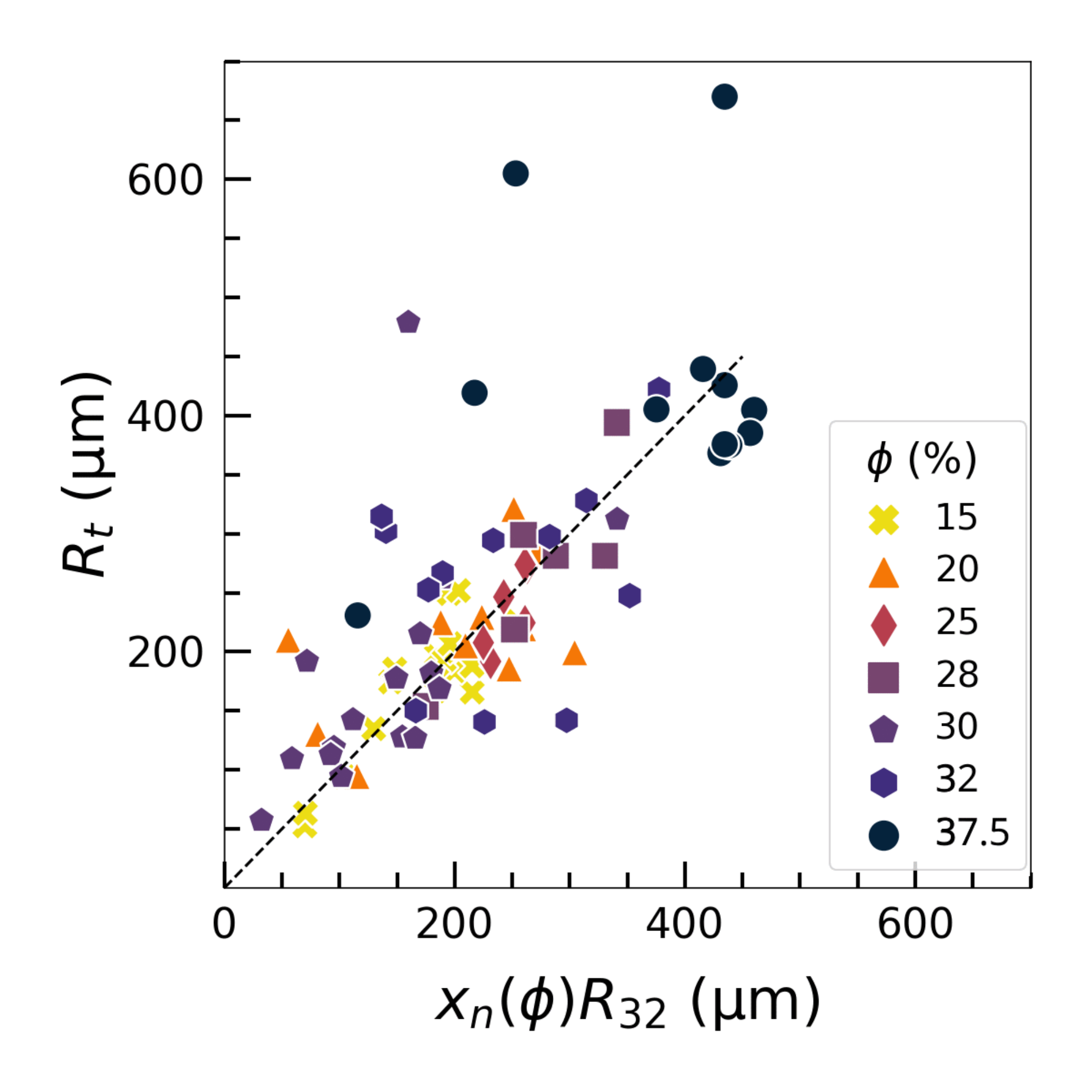}
    \caption{(left) Bubble radius $R_t$ as a function of the Sauter mean radius $R_{32}$, measured at the instant of the transition where the shrinkage rate slows down and the bubble starts to roam. 
    A linear relation (dashed line) is fitted to the data for each liquid fraction $\phi$, to determine the coefficient x$_n(\phi)$ defined in Eq.~3. 
    (right) Master curve of the bubble radius $R_t$ plotted as a function of  $x_n(\phi)\;R_{32} $. The dotted line has a slope equal to unity.}
    \label{fig:xn_liquid}
\end{figure}

\begin{figure}
    \centering
    \includegraphics[width=0.52\linewidth]{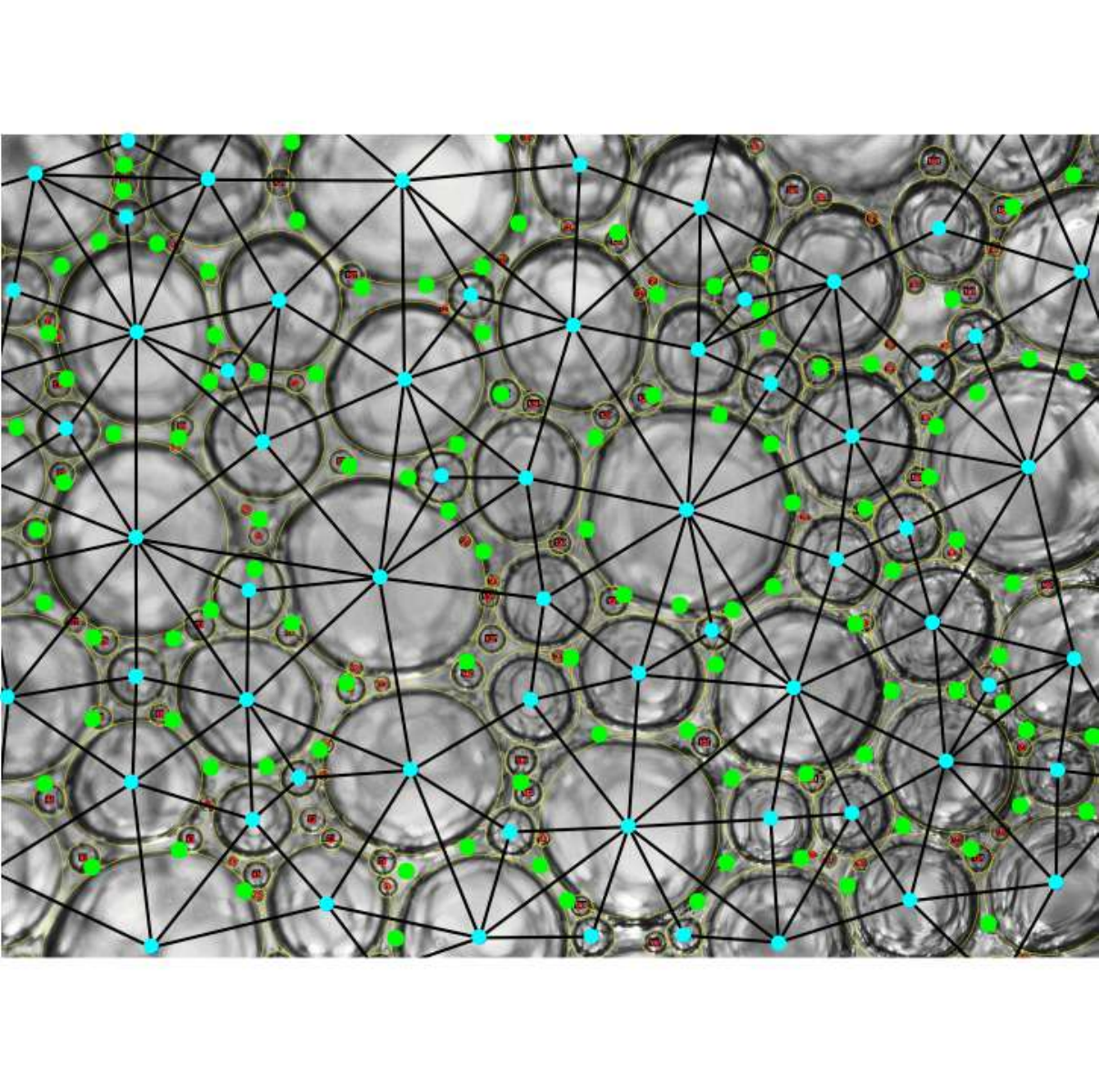}
    \includegraphics[width=0.47\linewidth]{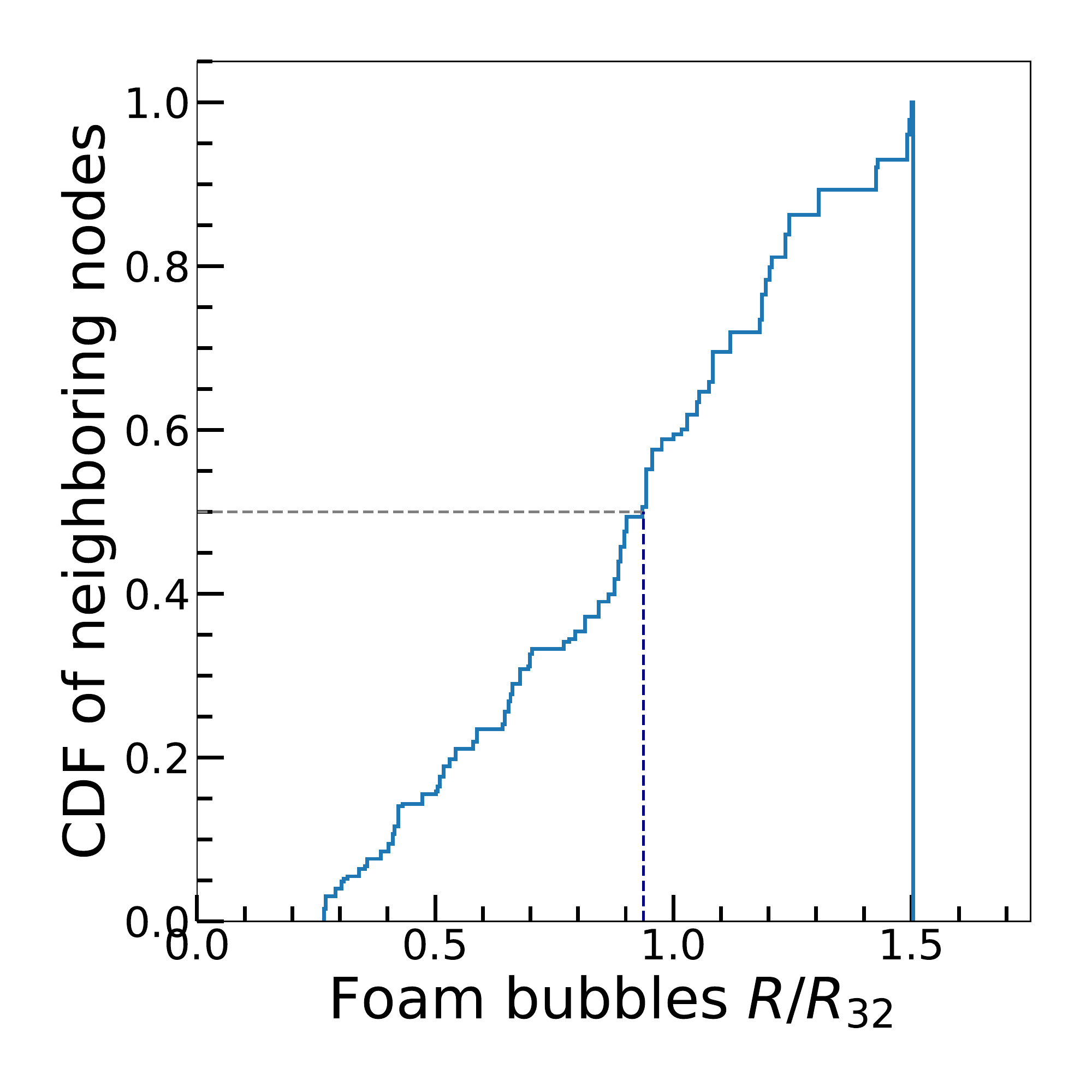}
    \caption{Image of a sample with $\phi = 15\%$ in the Scaling State (foam age $t= 9645 s$). The maximum radius of a roaming bubble, $R_t$, at this foam age is predicted from the value of $R_{32}$ at this age using Eq.~3. Each bubble is then classified as a roaming (resp. foam) bubble if its radius is smaller (resp. bigger) than $R_t$. In the figure, roaming (resp foam) bubbles are identified by  red (resp. cyan) spots at their centers.  We used the jammed bubbles' center positions to analyze the foam structure by triangulation (black lines joining the centers in overlay). We estimated the number of surface nodes, identified by green spots, and localized at the barycenter of the triangles. On average, we counted 1.5 roaming bubbles per node, and 1.2 nodes per jammed bubble. (right) CDF of the number fraction of nodes around bubbles as a function of their normalized radius $R/R_{32}$, evaluated for the sample shown in the left. We see that the median is very close to $R=R_{32}$ which means that nearly half of the nodes are delimited by a jammed bubble larger than $R_{32}$. This is consistent with the choice of $R_{32}$ as the characteristic radius of jammed bubbles constituting the nodes. Similar findings are found for liquid fractions up to $\phi_{rcp} \approx 31\%$.}
    \label{fig:network}
\end{figure}

\begin{figure}
    \centering
    \includegraphics[width=0.9\linewidth]
    {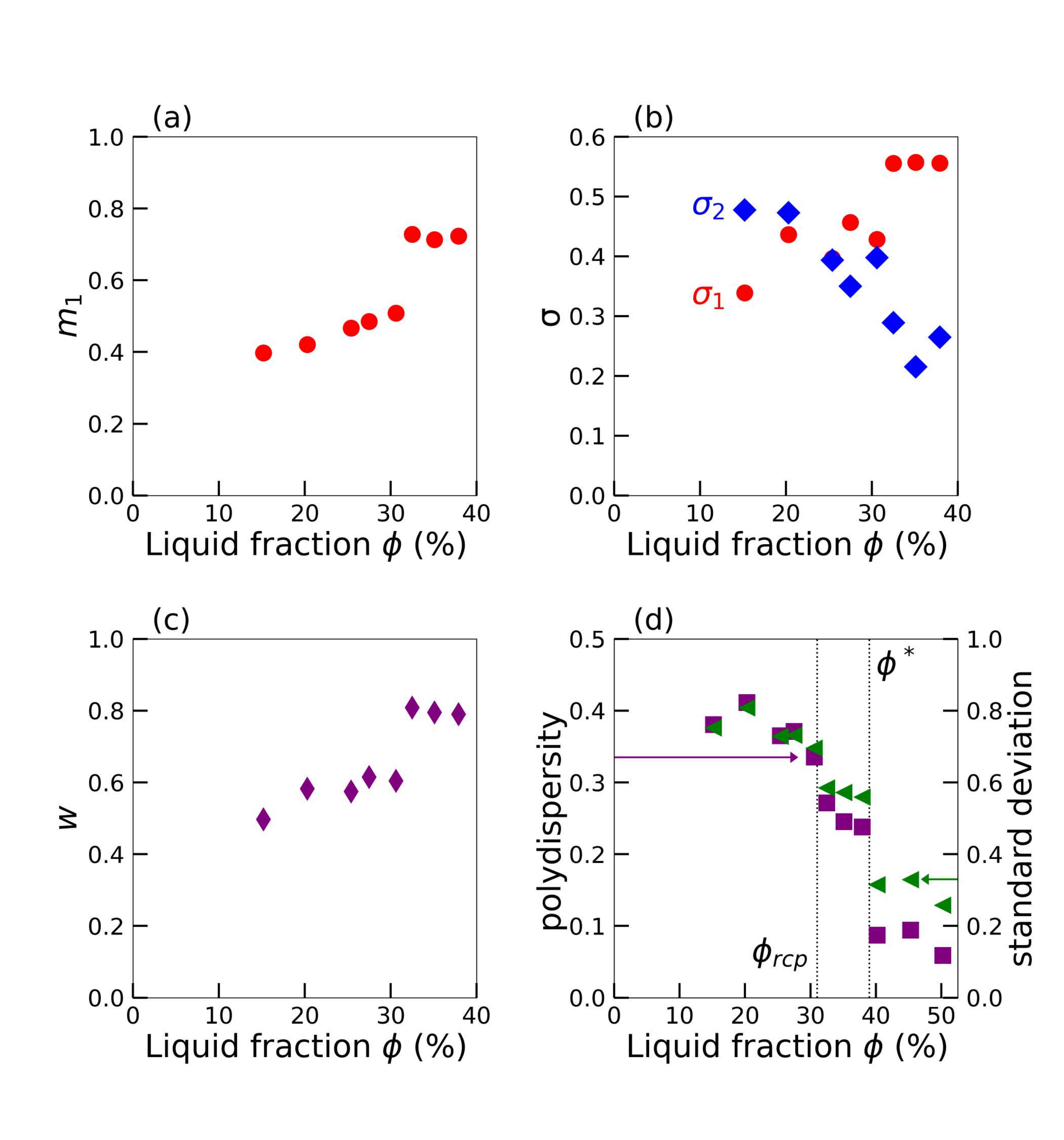}
    \caption{Variation of the fitted PDF parameters, as defined in the text (Eq.~6 and 7), with liquid fraction. a) Natural-scale median $m_1$ of the roaming bubbles (the median $m_2=1.66$ is fixed).  b) Logarithmic-scale standard deviation : $\sigma_1$ for the roaming bubble PDFs (red disks), $\sigma_2$ for the jammed bubble PDFs (blue diamonds). c) Relative weight $w$ of the roaming bubble distribution.  d) Polydispersity (squares), defined as ${R_{32}}/{\langle R^3 \rangle ^{1/3}}-1 $, and standard deviation (triangles) evaluated from raw data.
    Error bars are of the size or smaller than the symbol size.} 
    \label{fig:Evolution of fitting parameters}
\end{figure}

\begin{figure}
    \centering
    \includegraphics[width=0.7\linewidth]
    {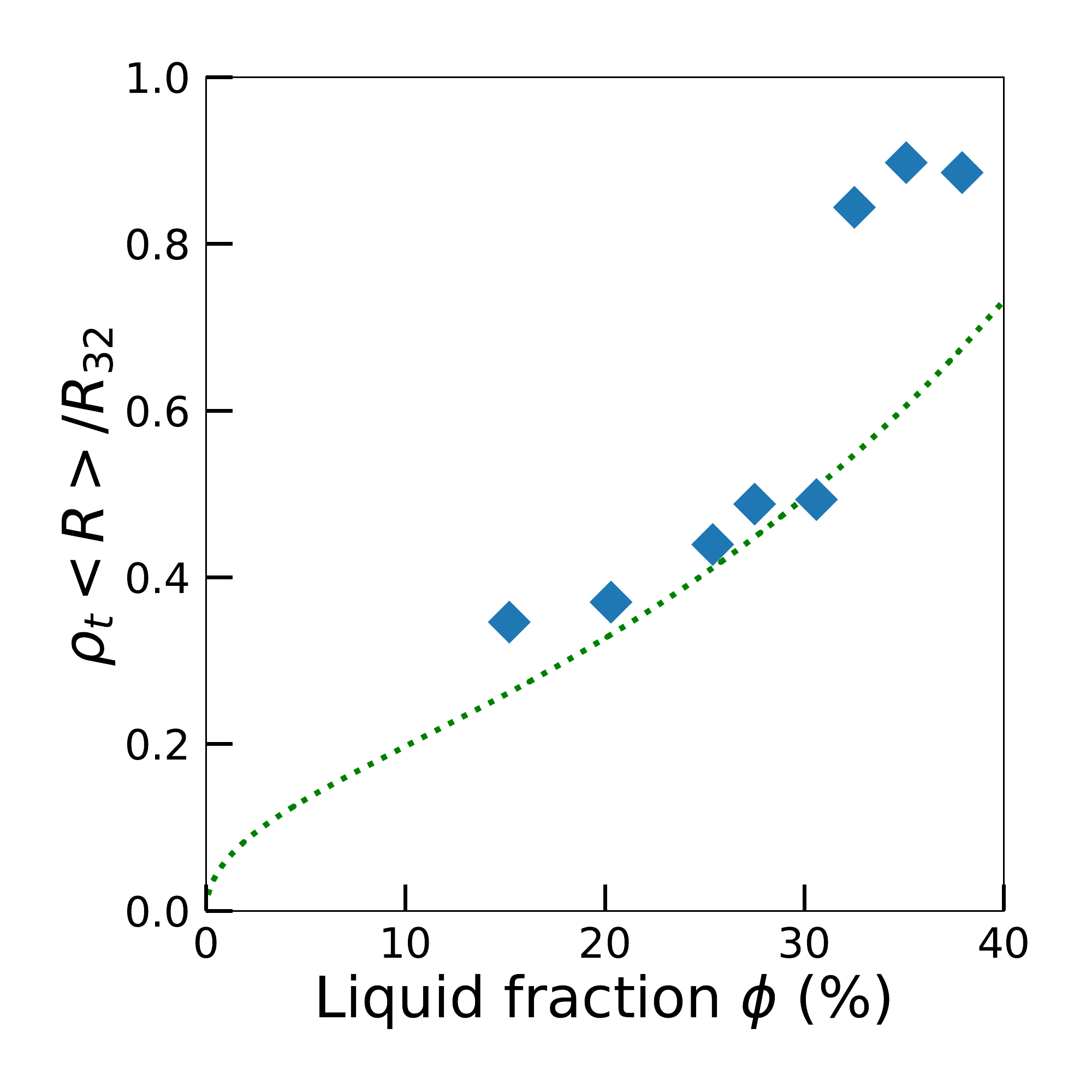}
    \caption{Characteristic width of the roaming bubble PDFs  $\rho_t$ (defined by Eq.~8) rescaled by $\langle R \rangle/R_{32}$ as a function of the liquid fraction. The dotted line represents Eq.~4 with $\xi = 2.2$ as in fig~2c. }
    \label{fig:Evolution of rho_t}
\end{figure}

\begin{figure}
    \centering
    \includegraphics[width=0.7\linewidth]
    {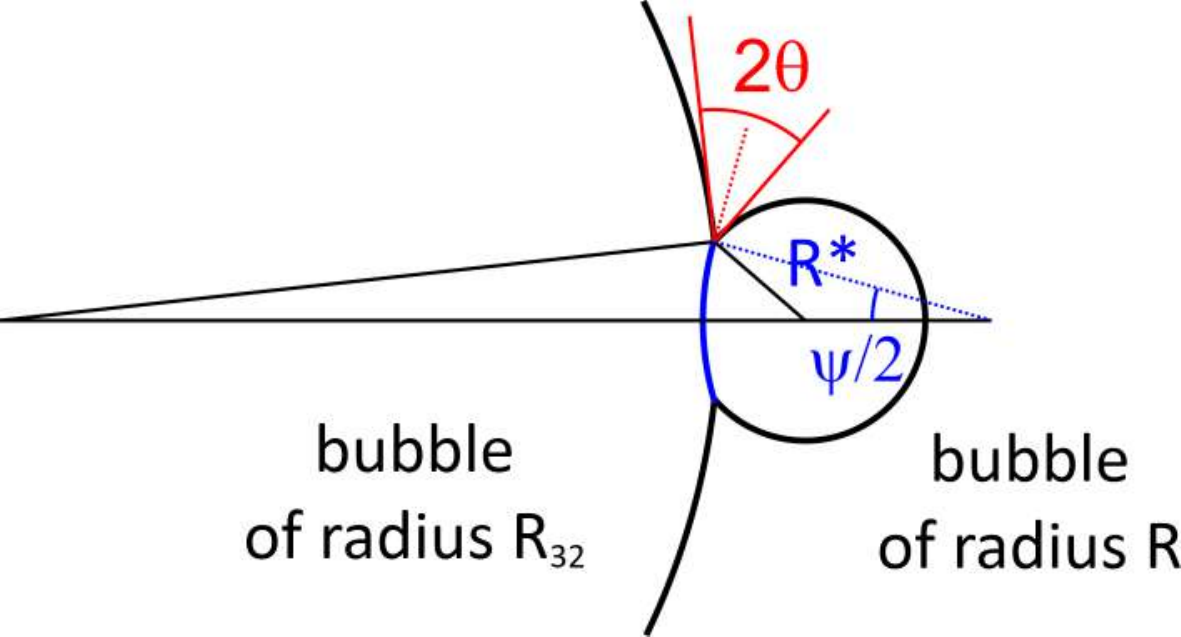}
    \caption{Bubbles of radii $R$ and $R_{32}$ sharing a film (in blue) due to adhesion forces accounted for by the contact angle $\theta$. $R^{\star}$ is the radius of curvature of the shared film.}
    \label{fig:adhesive_bubbles}
\end{figure}


\end{document}